\RequirePackage{fix-cm}
\documentclass[fleqn, final, smallcondensed,natbib,10pt]{svjour3}
\smartqed  
\usepackage{graphicx}
\usepackage{hyperref}
\usepackage{tabularx}
\usepackage{multirow}
\usepackage{mathtools}
\usepackage[dvipsnames]{xcolor}
\usepackage{colortbl}
\usepackage{booktabs}

\usepackage{rotating}
\usepackage{adjustbox}

\begin{document}

\title{PHANTOM: Curating GitHub for engineered software projects using time-series clustering\thanks{\textcolor{red}{Authors' Pre-Print submitted to Empirical Software Engineering Journal --- the final paper is available (Open Access) at \url{http://dx.doi.org/10.1007/s10664-020-09825-8}}}
}


\author{Peter Pickerill  \and
        Heiko Joshua Jungen \and
        Miros{\l}aw Ochodek \and
        Micha{\l} Ma{\'c}kowiak \and
        Miroslaw Staron
}


\institute{Peter Pickerill \at
             Department of Computer Science and Engineering, Chalmers University of Technology \\
              \email{peterpi@student.chalmers.se}           
           \and
           Heiko Joshua Jungen \at
           Department of Computer Science and Engineering, Chalmers University of Technology \\
           \email{jungen@student.chalmers.se}
           \and
           Miros{\l}aw Ochodek \at
           Poznan University of Technology \\
           \email{miroslaw.ochodek@cs.put.poznan.pl}
           \and
           Micha{\l} Ma{\'c}kowiak \at
           Poznan University of Technology \\
           \email{michal.mackowiak@cs.put.poznan.pl}
           \and
           Miroslaw Staron \at
           Chalmers University of Technology $|$ University of Gothenburg \\
           \email{Miroslaw.Staron@cse.gu.se}
}

\date{Received: date / Accepted: date}

\maketitle

\begin{abstract}


\noindent Context: Within the field of Mining Software Repositories, there are numerous methods employed to filter datasets in order to avoid analysing low-quality projects.  Unfortunately, the existing filtering methods have not kept up with the growth of existing data sources, such as GitHub, and researchers often rely on quick and dirty techniques to curate datasets. 

\noindent Objective: The objective of this study is to develop a method capable of filtering large quantities of software projects in a resource-efficient way. 

\noindent Method: This study follows the Design Science Research (DSR) methodology. The proposed method, PHANTOM, extracts five measures from Git logs. Each measure is transformed into a time-series, which is represented as a feature vector for clustering using the k-means algorithm. 

\noindent Results: Using the ground truth from a previous study, PHANTOM was shown to be able to rediscover the ground truth on the training dataset, and was able to identify ``engineered'' projects with up to 0.87 Precision and 0.94 Recall on the validation dataset.  PHANTOM downloaded and processed the metadata of 1,786,601 GitHub repositories in 21.5 days using a single personal computer, which is over 33\% faster than the previous study which used a computer cluster of 200 nodes. The possibility of applying the method outside of the open-source community was investigated by curating 100 repositories owned by two companies.

\noindent Conclusions: It is possible to use an unsupervised approach to identify engineered projects. PHANTOM was shown to be competitive compared to the existing supervised approaches while reducing the hardware requirements by two orders of magnitude.

\end{abstract}

\section{Introduction}
\label{sec:introduction}


Software project analysis used to be performed on small corpora of projects, as seen in industry-based case studies (\cite{Feldt2013}, \cite{Staron2013}, \cite{Staron2013a}). However, since the widespread use of code sharing sites such as GitHub\footnote{\url{https://github.com}}, researchers now have access to a massive corpus of software projects that can be analysed. Consequently, we see more and more studies in the area of software quality that base their research on mining GitHub repositories. Unfortunately, the quality of these repositories is unclear. It has been shown that most repositories can be considered to be of low quality, and could therefore skew analysis (\cite{Kalliamvakou2016}).

For instance, in the recent editions of the Mining Software Repositories (MSR) conference, a number of studies performed analyses based on GitHub repositories. The reported dataset sizes ranged between one to over 80,000 repositories (e.g. \cite{Schermann2017}, \cite{Gonzalez}, \cite{Noten2017}, \cite{Sadat2017}, \cite{Zhu2017}, \cite{Rausch2017}, \cite{Macho2017}), which is just 0.08\% of the current number. Most of these studies needed to filter out low-quality repositories from the collection, where low-quality denotes those repositories that do not fit into the desired sample. One of the frequently used approaches to filter such repositories is to rely on the project popularity. Unfortunately, it has been shown to perform poorly (\cite{Munaiah2017}). It is clear then, that for researchers to make use of the large number of repositories available on GitHub, new filtering methods are required. 

\cite{Munaiah2017} proposed a filtering method that outperformed traditional filtering approaches by using supervised classification. With this framework, over 1.8 Million GitHub repositories have been analysed, what makes it one of the largest datasets (\cite{Cosentino2017}). While this is an impressive achievement, a number of key issues with the method remain. First, the method analyses multiple artefacts to calculate the necessary metrics, i.e. Git logs, configuration files, source code, GitHub issues, license, unit tests. Consequently, the method has high computing-resource demands (in the study, a computer cluster with over 200 nodes to process the largest dataset and the task still required a month to complete). Second, some artefacts require dedicated processing, depending on the programming language and libraries used in the project. As a result, the proposed tool might require further work to adapt it to changing trends in programming technologies. Finally, the method requires multiple project artefacts to be downloaded and analysed; some of them might be out of interest for many studies using the method to pre-process the data.


With this in mind, in this paper, we propose a new method called PHANTOM\footnote{PHANTOM --- \url{https://github.com/Ionman64/PHANTOM}.} (Project History Analysis of Time-Series Method) that has the same purpose as the method proposed by \citeauthor{Munaiah2017}. It \emph{uses unsupervised learning to distinguish between ``engineered'' and ``not engineered'' projects at very large scale, solely based on their development history, while using commodity hardware.} We argue that it is possible to achieve comparable results to \cite{Munaiah2017} (which we refer to as the \emph{baseline study}) by applying the proposed unsupervised-learning method to filter repositories. Since the proposed method is unsupervised, it could also help the researchers to establish a ground truth by automatically grouping projects which have similar characteristics. By using the development history exclusively, we simplify the acquisition of data used for analysis, making it possible to perform the analysis on commodity hardware.  We validate PHANTOM on the datasets published in the baseline study (\cite{Munaiah2017}), extending it with additional 200 repositories. The method's efficacy is shown by using this dataset consisting of 850 labelled repositories from open source and 100 additional repositories owned by companies. PHANTOM's applicability for large-scale analyses is shown when it is applied to the dataset of over 1.8 Million software repositories. 

The structure of this paper is as follows. Section \ref{sec:related_work} presents and discusses related studies. The problem statement and the design of the validation study are presented in Section \ref{sec:methodology}. Section \ref{sec:phantom} describes the proposed method, which is further validated in Section \ref{sec:validation}. We discuss the findings from the validation study in Section \ref{sec:discussion}. Finally, we conclude our findings in Section \ref{sec:conclusions}.

\section{Related Work}
\label{sec:related_work}


The large-scale projects analyses (including code analysis) are mainly performed on open-source code repositories. With over 100 million Git repositories available, GitHub is becoming one of the most important sources data that could be used to study software source code and project-related phenomena. Unfortunately, the quality of the data available on GitHub is questionable. For instance, \cite{Kalliamvakou2016} performed the quantitative analysis of project metadata and manual investigation of a sample of 434 projects stored on Github to learn that the assumption that every repository available on GitHub contains a software project does not hold. In the studied sample, only 63.4\% of the repositories were related to software development. Also, most of the projects were personal---67\% of the projects had only 1 committer while 87\% had 2 or less. Finally, most repositories remained inactive or showed low activity (only 25\% of the studied repositories were active for over 100 days). Therefore, there is a need for filtering repositories that contain engineered software projects in order to reduce the chance of biasing the results of MSR studies by poorly engineered projects.

The simplest and often used approach to filter unwanted repositories is to rely on projects popularity, e.g. GitHub Stargazers (e.g. \cite{padhye2014study, ray2014large, casalnuovo2015assert, silva2016we, russell2018large}). The rationale behind using popularity as a filtering criterion is that it is assumed to be positively correlated with quality. However, as \cite{sajnani2014popularity} have shown, this might not be true.

Many studies used multi-stage pipelines to select a desirable sample of projects from GitHub. Most of them query publicly available services such as GHTorrent.\footnote{\cite{Gousi13}, \url{http://ghtorrent.org}} For instance, \cite{vasilescu2015quality, yu2015wait} selected projects by sequentially querying GHTorrent and Travis API in a funnel-like manner to study continuous integration practices. A similar approach was used by \cite{gharehyazie2018cross} who studied software clones. 

Some studies go beyond querying the metadata collected by GHTorrent by processing project repositories and extracting metadata from their artefacts. Good examples of studies implementing project analysers capable of processing such repositories are \cite{Gabel2010} and \cite{hebig2016quest}. They feature a high level of automation, yet the resources required (e.g. time, computing power) are large. \cite{Gabel2010} conducted a study about the uniqueness of source code within C++, C\# and Java applications taken from SourceForge.\footnote{\url{http://www.sourceforge.com}} They approached the question of ``how unique is software?'' by performing lexical analysis on a dataset of 6000 projects (in total 420 million lines of code). Following this, the percentage of unique code within 30 selected projects was measured. The study identified a general lack of uniqueness within software, where most programs are made from code snippets found in other software. The analysis time took four months, where source code was compared on a token-level with an optimised tool. This shows that source-code analysis requires a lot of time even for a small number of projects. Similarly, in the study by \cite{hebig2016quest} that aimed to find GitHub repositories containing UML models, the process of downloading and processing the repositories took 6 weeks.

\cite{Cosentino2017} looked into 80 studies that mined GitHub. They found that the two largest data sources, GHTorrent and GitHub API\footnote{\url{https://developer.github.com/v3}}, were criticised by researchers. The GitHub API was said to be a source of problems, given request limitations and errors in the data returned. They state that \emph{``the GitHub API request limit acts as a barrier to getting data from GitHub''},  which affects curated datasets (such as GHTorrent) and individual researchers. Many researchers criticise the size and up-to-dateness of services like GHTorrent. \citeauthor{Cosentino2017} report that of studies they explored, only three looked at more than 100,000 repositories. These findings show that despite using these services, researchers struggle to collect up-to-date data at large scale.

\cite{Robles2017} identified similar issues with GitHub when collecting information about twelve million repositories. First, due to API limits, they calculated that it would have taken fourteen months to collect all of the data using a single API key, adding that \emph{``[\ldots] this would have made the data gathering unfeasible.''} To gather the data in a feasible time, twenty keys were used. Secondly, 25\% of the twelve million repositories had been moved or deleted between the time GHTorrent collected its data and the researchers request to the GitHub API, which wasted keys and analysis time. This shows that data collection time can be reduced by using volunteered keys for data collection, a practice employed by GitHub mirroring sites, like GHTorrent and Boa.\footnote{\cite{dyer2013boa}, \url{http://boa.cs.iastate.edu}}

\cite{Kalliamvakou2016} observed that there are different ways to merge commits and GitHub cannot detect all of them. Therefore, some merges are not reported through the API. A further peril when using the GitHub API is that, unlike cloning with Git, the GitHub API does not redirect requests when a repository was moved. Accessing a moved repository with the API will result with a not found status code.

\cite{Nunez-Varela2017} conducted a systematic review of 226 papers studying source-code metrics. They identified that most studies considered one programming language and paradigm. Over 85\% of the studies use object-oriented metrics. This is reflected in the available public datasets and metric-extraction tools. While the paper does not reason why researchers focus on one language, it indicates that cross-language analysis of source code may not be straightforward. They claim that although the use of metrics \emph{``theoretically can be applied to any language''}, in practice, it is complex and tools do not support all languages.

Recently, \cite{Munaiah2017} proposed a new automatic method along with a tool called \textit{reaper} that can be used to curate GitHub for engineered software projects. The method classifies GitHub repositories using seven dimensions: Community, Continuous integration, Documentation, History, Issues, License, and  Unit testing. \textit{reaper} was used to download and process 1.8 million GitHub repositories. The analysis required a computer cluster with over 200 nodes and took over a month to complete. In the next step, the authors sampled the collected dataset and manually labelled 500 repositories, of which 200 were used for validation. In addition, they analyzed and manually labelled 150 repositories owned by organizations such as Amazon, Apache, Facebook, Google, and Microsoft. They trained and validated four classifiers (Random Forest and custom score-based classifier) using the collected data and applied them to predict the number of repositories hosted on GitHub that contain engineered software projects. Depending on the classification algorithm and training dataset, the predictions of the number of such repositories ranged between 6\% and 70\%. The most accurate classifier predicted that only around 24\% of the repositories contained engineered projects. Also, they showed that the proposed classifiers outperform filtering repositories by their popularity. 

Our study proposes a light-weight method for filtering projects that complements the methods proposed in the above studies. The method processes only one project artefact, the Git log, and extracts five measures directly from it (Integration Frequency, Commit Frequency, Integrator Frequency, Committer Frequency, and Merge Frequency). Each of the measures (being a time series) are characterised by a set of over forty manually-designed features (e.g. duration, amplitude, positive and negative gradients). Since the method directly analyses the Git logs, it does not suffer from the limitations of GitHub API or GHTorrent. Secondly, it is programming language agnostic because it does not analyse the source code. Its closest counterpart is the filtering method proposed by \cite{Munaiah2017}, since any both methods are based on machine-learning algorithms and have the purpose of filtering engineered software projects by a defined quality. The main difference between the two approaches is that they use different artefacts to extract information about the projects and different machine-learning algorithms. The approaches to filtering projects proposed by other authors were developed to select projects having particular, explicitly known characteristics that were interesting for a given study (e.g. \cite{vasilescu2015quality} selected projects not being forks, written in certain programming languages, and having 200+ pull requests). In the case of distinguishing between engineered not engineered projects, it is difficult to explicitly provide such characteristics. However, even for such studies and others that require processing other project artefacts than Git logs, the proposed method could be used as a scanning tool as the first step to quickly identify and reject repositories containing not engineered projects and prevent wasting resources to download and process their artefacts.

\section{Research Methodology}
\label{sec:methodology}

The research conducted in this paper follows the Design Science Research (DSR) methodology (\cite{hevner2004design}). In particular, we decided to follow the guidelines provided by \cite{Wieringa2014}. 

DSR is a problem-solving paradigm that focuses on creating and evaluating artefacts and solutions for practical purposes.  In design science, research follows the engineering cycle---it is conducted iteratively until the objectives are reached. The engineering cycle consists of five steps:
\begin{enumerate}
    \item Problem investigation --- understanding the problem and outlining the steps to solve the problem and evaluate the solution. 
    \item Treatment design -- designing an artefact which is used in the study (e.g. a computer program).
    \item Treatment validation --- evaluating the artefact in a context similar to the context where it is to be used (e.g. lab environment). 
    \item Treatment implementation --- introducing the artefact into the context where it is to be used (e.g. software development organization).
    \item Implementation evaluation --- evaluation of the effects of the introduction of the artefact on the context (e.g. check whether the program improved the process of software development).   
\end{enumerate}

The first three steps of the engineering cycle are called the design cycle.  According to \citeauthor{Wieringa2014}, the design cycle is what is usually performed by researchers when designing an artefact while the remaining two steps (treatment implementation and evaluation) can be done once the artefact is in the hands of its intended users. 

The design cycle begins with an investigation of the problem to gain an in-depth understanding of the causes. The acquired knowledge is then used to design a treatment. Treatment is defined as the interaction of an artefact with a problem context.  The goal of the third step, treatment validation, is to confirm that the designed treatment satisfies all the requirements and whether the treatment is able to treat the problem.

In our study, we performed a full design cycle to create a new method---PHANTOM which is the method to analyse software projects, and the problem context is to filter engineered projects from GitHub.  Firstly, we investigate the problem and define the requirements for the method by analysing the limitations of the \emph{reaper} tool proposed in the study by \cite{Munaiah2017}.  Then we perform a validation of the method by applying it to the datasets provided in the baseline study and comparing the results between PHANTOM and \emph{reaper}. 

The goal of two remaining steps in the engineering cycle is to apply the treatment to and investigate how it interacts with their real-world context. Treatment evaluation is a scalable process and usually requires multiple applications of the treatment to get a full understanding of its usefulness. Although we do not aim to evaluate the method in this study, we apply  PHANTOM to replicate the task of filtering nearly 1.8 Million real software repositories stored on GitHub in the baseline study, what could be perceived as an initial evaluation of the method in its real-life context, and apply it to recognise engineered projects in industrial Git repositories.

In the remaining part of this section, we investigate the problem of filtering software repositories that justifies the need for designing PHANTOM. Then, we discuss the validation and evaluation procedures used in this study.

\subsection{Problem Investigation}
\label{sec:problem_investigation}

 \cite{Munaiah2017} provides an abstract definition of an engineered software project which states that \emph{``a software project that leverages sound software engineering practices in one or more of its dimensions such as documentation, testing, and project management.''}  Later in their study, they customise it by stating that \emph{``(a) an engineered software project is similar to the projects contained within repositories owned by popular software engineering organizations such as Amazon, Apache, Microsoft and Mozilla and\footnote{Although the definition provided by  \cite{Munaiah2017} uses \emph{and} as the conjunction between parts (a) and (b), we believe it should be \emph{or} since they exclusively train classifiers either on the repositories owned by the organisations or on the repositories containing general-purpose utility projects.} (b) and engineered software project is similar to the projects that have a general-purpose utility to users other than the developers themselves.''} Since we partially base our study on the dataset provided by \cite{Munaiah2017}, we follow these definitions of an engineered software project. 
 
The existing methods for filtering projects (and in particular filtering engineered projects) make trade-offs between the depth of analysis and performance. The lightweight approaches use community-based filtering strategies, for example; popularity, issues, and forks are common but do not measure the internal quality of a project. Unfortunately, these approaches are typically less accurate than those relying on analysing project artefacts \citep{Munaiah2017} and \emph{cannot be used on repositories without community interaction}, for instance, those that are hosted on private servers. The methods that perform deeper analyses of project artefacts (e.g. \emph{reaper} proposed by \cite{Munaiah2017}) are more accurate but require high computing power. These methods often analyse multiple project artefacts, including source code, which is time-consuming and require implementing and maintaining dedicated analysis tools (e.g. to analyse code written in different programming languages). To mitigate performance problems, these methods often rely on the metadata collected by publicly available services such as GHTorrent. Unfortunately, multiple studies have reported problems and limitations of this approach. Also, the existing methods perform filtering based on the cross-sectional assessments of projects, i.e. without taking into account their history. Finally, the methods that base filtering on supervised machine-learning algorithms require manual labelling of the training data which is time-consuming. 

Therefore, it seems that if one wants to create a new filtering method that analyses project artefacts, they should focus on \emph{reducing hardware requirements}. Also, it is worth considering basing the filtering on \emph{chronological measures}, which not only reflect the current status of the project but also its history. Finally, it would be preferable if the method \emph{does not require manual labelling of data}, and instead use unsupervised methods, such as clustering.

Chronological measures taken from repositories can be represented as a time-series. In the field of time-series clustering, a number of techniques are available. Euclidean Distance and DTW \citep{Ratanamahatana2004} are most commonly used, however they cannot compare time-series of very different lengths. Feature-based approaches have been shown as an effective way to compare time-series of different lengths and reduce the dimensionality of the data significantly (e.g. \cite{Wang2006}, \cite{Deng2013}, \cite{Fulcher2014}, \cite{Guo2008}). However, extracted features must describe the time-series correctly to ensure accurate clustering. 

Git logs seem to be one of the most frequently analysed artefacts to extract features used for filtering projects. Also, it allows extracting chronological measures describing trends in the development of a software project (e.g. commit frequency, number of contributors). Commit frequency can tell much about the process and practices employed by a software development team and reflects the changes made in the ways of working \citep{zhao2017impact}.  Also, the frequency of commits correlates with the number of bug-introducing changes in software \citep{eyolfson2014correlations}.  The relationship between commit frequency and overall quality of a project is also observed by \citep{kolassa2013empirical}, who stated that ``the commit frequency is a fast indicator to determine if the project is healthy because it has regular contributions and if the developers are productive by checking whether they contribute regularly.'' \cite{kalliamvakou2015open} observed that industrial projects have more merges resulting from pull requests.  Even the number of reverted commits could be an indicator of how the project team operates. According to \cite{shimagaki2016commits} many reverted commits could be avoided if a team has good communication practices and high change awareness.   Therefore, it is reasonable to assume that the information about commit frequency might be a useful source of information when filtering engineered projects. We could expect that engineered projects will exhibit different characteristics of commit frequency in time (including merging commits) not present in not engineered projects.

Taking the above into account, we define the following requirements for a new project-filtering method, which we call PHANTOM:

\begin{itemize}
    
    \item \emph{Req1}: \emph{All measures must be extractable from a Git log.} --- we want to limit the number of artefacts that need to be processed to a single one, being Git log.
    
    \item \emph{Req2}: \emph{Time-series must be characterised by feature vectors accurately.} --- we want to extract chronological measures from Git log; since the feature-based clustering seems to be the best option to compare time-series of different lengths, we need to find a set of features that will correctly capture the most important characteristics of the time-series \citep{Esling2012}. 
    
    \item \emph{Req3}: \emph{The established ground truth can be discovered using unsupervised learning (without knowing the decision classes).} --- we have to verify that the proposed method is able to identify clusters that are meaningful from the perspective of the problem of filtering repositories containing engineered software projects.
    
    \item \emph{Req4}: \emph{The method performs well on commodity hardware at large-scale.} --- it is the key requirement; it addresses the most important limitation of the existing filtering methods that analyse project artefacts.
    
    \item \emph{Req5}: \emph{The method provides comparable accuracy to supervised methods.} --- the existing approaches make a trade-off between accuracy and performance; the proposed method is supposed to provide similar accuracy while reducing the performance requirements.
    
    \item \emph{Req6}: \emph{The method can be used to filter projects in different contexts.} --- there is a threat that the performance of an unsupervised method can be specific to a single dataset only; therefore, we need to perform studies on different datasets coming from both open-source community and industry to mitigate that risk.
    
\end{itemize}

\subsection{Treatment Validation}
\label{sec:val-eval-methodology}

In order to validate PHANTOM, we performed a series of simulation studies on the datasets published in the baseline study. 

\cite{Munaiah2017} published five datasets. These datasets are formatted as collections of GitHub repository URLs. Four of these datasets (Organization\footnote{We preserve the original spelling of the dataset name used in the study by \cite{Munaiah2017}.}, Utility, Negative Instances, and Validation) are used as ground truth in the validation of the proposed method, with the fifth being referred to as the Large dataset (a collection of over 1.85 Million URLs). 

To create the ground truth datasets, \citeauthor{Munaiah2017} followed a manual curation process in order to label repositories. Each repository was independently judged by two or three researchers as either engineered or not, according to agreed guidelines. If the judgement about a repository differed, it was either discussed further or discarded. The datasets are summarised in the list below:

\begin{itemize}
\item \textbf{Organization} --- it consists of a set of 150 engineered repositories. Engineered projects are defined as similar to those of popular software engineering companies such as Amazon, Apache and Facebook. The researchers manually investigated repositories to find those project that matched the definition.

\item \textbf{Utility} --- it consists of another set of 150 engineered repositories. It defines an engineered project as one with a general-purpose. That is to say, a repository that has value to users other than the developers. The repositories were randomly sampled from 1,857,423 GitHub repositories.

\item \textbf{Negative Instances} --- it holds 150 repositories that are not engineered. The repositories do not conform to either of the definitions of engineered project. The dataset resulted from the selection process of the Utility dataset, which means that it contains the first 150 repositories that both authors rejected.

\item \textbf{Validation} --- it consists of 100 engineered and 100 not engineered project repositories. The selection process is similar to the one of the Utility dataset and shares the definition of what is engineered and not. 

\item \textbf{Large dataset} --- it is  a collection of 1,857,423 GitHub URLs. In contrast to the other datasets, there is no ground truth, meaning the quality of the repositories is unknown. 

\end{itemize}

Since the four ground-truth datasets contained labelled data, they allowed us to validate the PHANTOM's accuracy and compare it with the accuracy of \emph{reaper} proposed in the baseline study. While evaluating the accuracy of the methods we used the popular prediction quality metrics Precision, Recall, F-Measure, and Matthews Correlation Coefficient (MCC).

When predicting the label of an object and comparing the predicted label to the actual label there are four possible outcomes: the object is correctly classified to a positive class---true positive (TP), the object is falsely classified to a positive class---false positive (FP), the object is correctly recognised as not belonging to the positive class---true negative (TN) and the object is falsely recognised as not belonging to the positive class---false negative (FN). We calculate  Precision, Recall, F-Measure, and Matthews Correlation Coefficient (MCC) by aggregating information about the outcomes for all classified cases. 

We extended the previous study by comparing the accuracy of the baseline and PHANTOM models on another sample of projects from the Large dataset. The goal of this analysis was two-fold. First, it allowed us to compare the accuracy of the models on a new dataset. Second, we were able to verify the validity of predictions regarding the number of engineered software projects hosted on GitHub reported in the baseline study and the ones obtained with the PHANTOM models. 

We classified all instances from the Large dataset using the baseline and PHANTOM models (we used a best-performing PHANTOM model for each of the measures considered by PHANTOM). Since we wanted to understand the similarities and differences between the prediction made by the models, we used stratified sampling to select 5 samples, each containing 50 project instances (the total number of 250 instances):
\begin{itemize}
    \item \emph{True/True} --- this was a sample of instances for which the best-performing baseline models and PHANTOM models trained on the Utility dataset\footnote{The baseline study concluded that the best-performing models were trained using the Utility dataset.} unanimously classified them as engineered projects. 
    
    \item \emph{False/False} --- we randomly selected instances that were unanimously\footnote{We ignored the predictions made by PHANTOM model trained using the 'merges' time-series after we had discovered that it was indicating nearly all repositories as engineered.} classified as not engineered projects. 
    
    \item \emph{False/True} --- a random sample of instances which were unanimously predicted to be not engineered projects by the PHANTOM models and unanimously predicted as engineered projects by the best-performing baseline models.
    
    \item \emph{True/False} --- a random sample of instances which were unanimously predicted to be engineered projects by the PHANTOM models and unanimously predicted as not engineered projects by the best-performing baseline models.
    
    \item \emph{Mixed} --- a random sample of the remaining repositories not fitting to any of the above categories. These were instances for which there was a partial agreement between the prediction models.
\end{itemize}

All of the selected instances were then independently labelled by two or three authors using the same criteria as in the baseline study for the Utility dataset. By using this dataset, we were able to analyse how much the baseline and PHANTOM models complement or contradict each other. Consequently, we were able to assess the impact of the differences in the way they classify repositories on their predictions of the number of engineered projects hosted on GitHub. 

Finally, we applied the best-performing PHANTOM models to filter industrial projects to investigate whether it is possible to use PHANTOM in different contexts. We introduced a new dataset (\textbf{Industry}) that contains 100 repositories belonging to two companies. Company A is one of the fastest-growing software agencies in the European Union. With its presence on the market for more than ten years, it has successfully delivered over 1400 projects. Currently, it employs more than 500 employees and develops products for multiple sectors, e.g. FinTech, Healthcare, Tourism, E-commerce, Entertainment, and e-Government. It maintains around 500 Git repositories in a private GitHub space. Company B develops embedded software for infrastructure projects. The dataset was labelled by the employees of the companies.

\section{PHANTOM---A Developed Artefact}
\label{sec:phantom}

\begin{figure}
    \includegraphics[width=\linewidth, keepaspectratio, page=1]{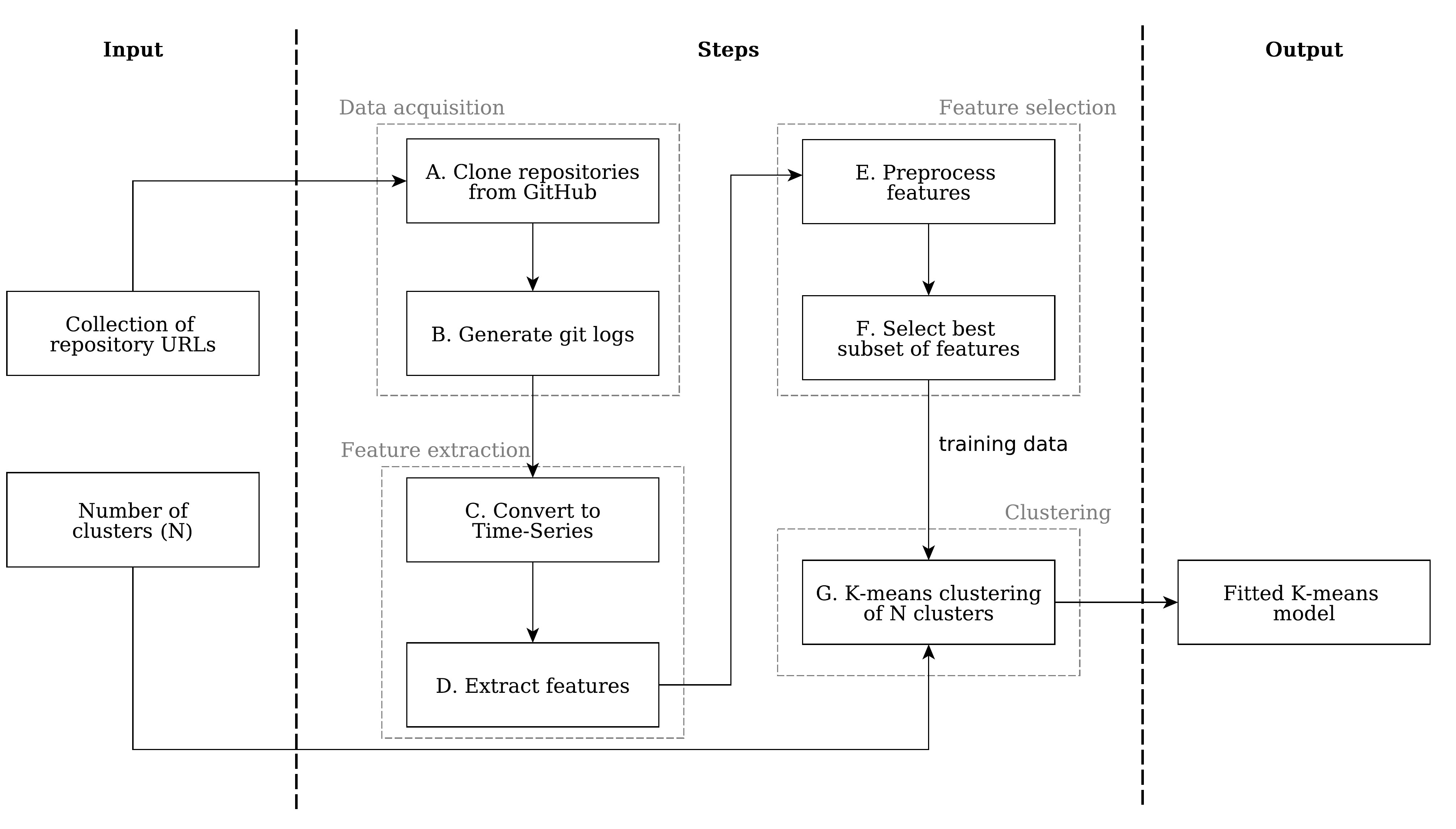}    
    \caption[Overview of PHANTOM]{Overview of PHANTOM.}
    \label{fig:phantom_overview}
\end{figure}

PHANTOM (Project History Analysis of Time-Series Method) is a software-repositories filtering method that addresses the problems identified in Section \ref{sec:problem_investigation}. The process used in PHANTOM is presented in Figure \ref{fig:phantom_overview} and its steps are explained in the subsequent paragraphs.

The input to the method is a collection of repository URLs, which locate the repositories to be analysed. Each repository is cloned to the machine using Git (Step A). Next, the Git log of the cloned repository is generated (Step B) in the format defined in Table \ref{table:log_format}, where each column is separated by a comma (“,”). From now on, Git log will refer to this format.

\begin{table}
    \centering
    \caption[Git Log Format]{The format of the generated Git logs, each column is separated with a comma (UTS refers to Unix Timestamp, GUID refers to Globally Unique Identifier).}
    \label{table:log_format}
    \begin{tabularx}{\linewidth}{l|XX|XXX|XXX}
    \hline\noalign{\smallskip}
         \multicolumn{1}{c}{}&
         \multicolumn{2}{c}{Hashes}    &
         \multicolumn{3}{c}{Author}        &
         \multicolumn{3}{c}{Committer}    \\
         
        Name&
        \multicolumn{1}{c}{Commit} & \multicolumn{1}{c|}{Parent}& 
        \multicolumn{1}{c}{Name} & \multicolumn{1}{c}{Email} & \multicolumn{1}{c|}{Date} & \multicolumn{1}{c}{Name} & \multicolumn{1}{c}{Email} & \multicolumn{1}{c}{Date} \\ 
        \noalign{\smallskip}\hline\noalign{\smallskip}
        Format     & \%H                        & \%P                        & \%an                     & \%ae                      & \%at                     & \%cn                     & \%ce                      & \%ct                     \\
        Type       & GUID        & GUID  & String            & String                    & UTS           & String                   & String                    & UTS \\
        \noalign{\smallskip}\hline
    \end{tabularx}
    
\end{table}

An example Git log is presented in Table \ref{table:log_example}. The Git log contains timestamped rows, which makes the conversion to time-series possible (Step C). These time-series use combinations of the different columns of the Git log and are referred to as measures. We extract five measures: Commit Frequency (Commits), Integration Frequency (Integrations), Commiter Frequency (Commiters), Integrator Frequency (Integrators), and Merge Frequency (Merges). These measures are defined in Table \ref{table:phantom_measures}. As previously mentioned in the problem investigation (Section \ref{sec:problem_investigation}), many studies observed that commit frequency could be a valuable source of information about the process and practices used in software development projects. Our hypothesis is that the measures we extract from Git logs allow separating engineered and not engineered projects, in a given dataset. They can be divided into two groups. Commits, Integrations, and Merges characterise ways of working and peace of the project. For instance, we could see periodic increases in the frequency of commits, or long periods of inactivity, etc. The frequency of commits seem to correlate with the quality of the project \citep{eyolfson2014correlations, kolassa2013empirical}, therefore, they could help discriminate between engineered and not engineered projects. Similarly, the code merging practices might be different depending on the capabilities of projects, as the number of merges tends to be greater in commercial projects \cite{kalliamvakou2015open}. The remaining two measures (Commiters and Integrators) provide information about the number of people involved in the project at a given period of time. We expect that more attractive projects bring more attention from the community and consequently more people contribute to their codebase.

\begin{table}
    \centering
     \caption[Example Git Log]{Example Git log, where each row is one commit.}
    \label{table:log_example}
    \begin{tabular}{ll|cll|cll}
    \hline\noalign{\smallskip}
    
    \multicolumn{2}{c|}{Hashes}                              & \multicolumn{3}{c|}{Author}                                                      & \multicolumn{3}{c}{Committer}                                                   \\
    {Commit} & {Parent} & {Name} & {Email} & {Date} & {Name} &{Email} & {Date} \\ 
    \noalign{\smallskip}\hline\noalign{\smallskip}
    b57f4f3                    & 82c9f95                    & ab                       & a@b.com                   & 1519904296               & cd                       & c@d.com                   & 1519904396               \\
    82c9f95                    & efaf9cd                    & ab                       & a@b.com                   & 1519834072               & ab                       & a@b.com                   & 1519904296               \\
    efaf9cd                    & 703b7b1                    & ab                       & a@b.com                   & 1519404672               & ab                       & a@b.com                   & 1519824672  \\             
    \noalign{\smallskip}\hline
    \end{tabular}%
   
\end{table}

\begin{table}
    \centering
    \caption[Measures extracted by PHANTOM]{The measures extracted by PHANTOM.}
    \label{table:phantom_measures}
    \begin{tabularx}{\linewidth}{p{2.5cm} | p{2.5cm} X}
    \hline\noalign{\smallskip}
    Measure  & Git Log Information Used  & Description\\ 
    \noalign{\smallskip}\hline\noalign{\smallskip}
    
    Commit~Frequency (Commits)
    & Author Date                           
    & The number of commits summed per Week                                           \\
    
    Integration~Frequency (Integrations)
    & Committer Date                        
    & The number of integrations summed per Week.                                     \\
    
    Committer~Frequency (Committers)        
    & Author Date, Author Email             
    & The number of unique developers (by email) that have made commits per Week      \\
    
    Integrator~Frequency (Integrators)
    & Integrations Date, Integrations Email 
    & The number of unique developers (by email) that have made integrations per Week \\
    
    Merge~Frequency (Merges)             
    & Parent Hashes, Committer Date         
    & The number of merges summed per Week    \\
   
    \noalign{\smallskip}\hline
    
    \end{tabularx}
    
\end{table}

In Table \ref{table:git_log_timeseries_example}, the example Git log is transformed into the five measures. Measures are represented as a regular time-series, which makes their comparison possible; however, considering very different lengths, a direct comparison of the time-series may not make much sense. Interpolation would mean that the data are manipulated, which the authors argue would not be a true representation of the development history. The time-series are of very different lengths (e.g. 50 weeks and 900 weeks). Therefore, DTW and Euclidean distance are not suitable. Instead, a feature-based approach is selected, which does not come with these issues.

\begin{table}
    \centering
    \caption[Example Time-Series for the Five Measures]{Example time-series for the five measures.}
    \label{table:git_log_timeseries_example}
    \begin{tabular}{l | ccccc}
        \hline\noalign{\smallskip}
        Date       & Integrations & Integrators & Commits & Authors & Merges \\
        \noalign{\smallskip}\hline\noalign{\smallskip}
        2018-02-26 & 2           & 2          & 2      & 1      & 0      \\
        2018-02-19 & 1           & 1          & 1      & 1      & 0     \\
        \noalign{\smallskip}\hline
    \end{tabular}
    
\end{table}

\begin{figure}
    \includegraphics[width=\linewidth,keepaspectratio]{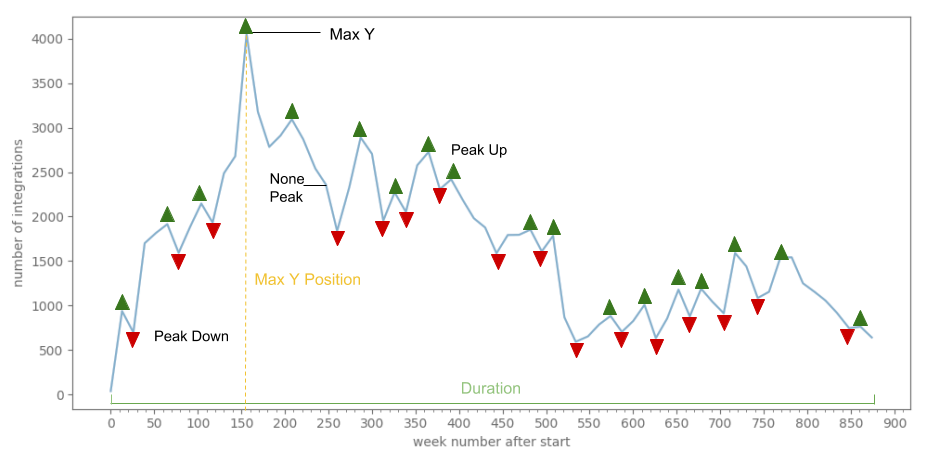}
    \caption[Time-Series Features Example 1]{Example features that can be extracted from a time-series.}
    \label{fig:time-series_features_example1}
\end{figure}

The time-series are therefore reduced in dimensionality by extracting a fixed-length feature vector (Step D). In the feature-based approach, a time-series is represented by a feature vector. This feature vector is fixed-length which makes it compatible with common clustering algorithms. A feature vector consists of a number of values that describe certain aspects of time series. For example, a feature could be the lowest value within the sequence, which could be called \textit{Min Y}. Features can show very simple, or very  complex characteristics (see Figure \ref{fig:time-series_features_example1}). The up and down peaks of the time-series are marked using upward and downward pointing triangles. To calculate the features \textit{Peak Up} and \textit{Peak Down} these points are counted. A peak can be described as any point that is either higher or lower than the preceding and succeeding points. \textit{Peak None} is the sum of all points that are not marked (e.g. at week 250). The feature \textit{Max Y} is the largest value within the time-series, which is roughly 4000 in the example. The position of \textit{Max Y} is captured by \textit{Max Y Pos}, which is the index (week) in which the value occurred, around 150 in the example. \textit{Duration} is equal to the total number of weeks between the first point and the last point, illustrated by the bar close the x-axis. 

\begin{figure}
    \includegraphics[width=\linewidth,keepaspectratio]{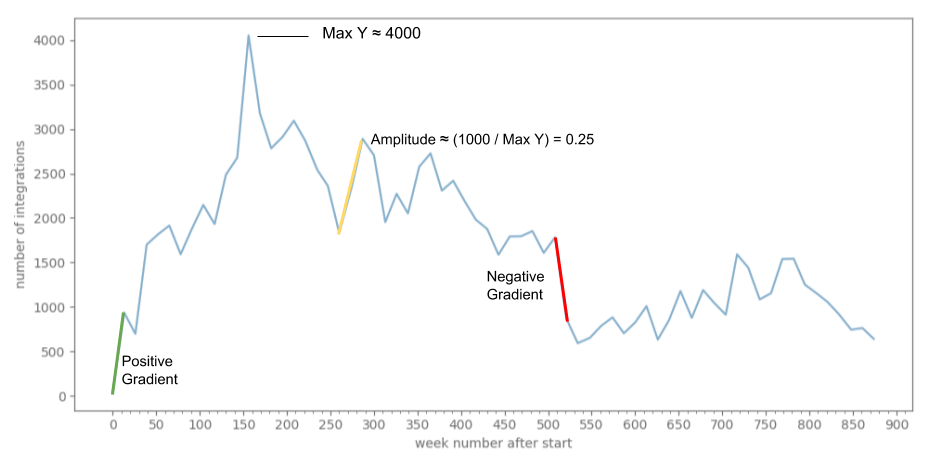}
    \caption[Time-Series Features Example 2]{Example features that can be extracted from a time-series.}
    \label{fig:time-series_features_example2}
\end{figure}

In Figure \ref{fig:time-series_features_example2} more features are illustrated. At the start of the time-series, a subsequence is labelled to show a positive gradient. All positive gradients between neighbouring observations are averaged to calculate the feature \textit{Mean Positive Gradient}. Similarly, the feature \textit{Mean Negative Gradient} is calculated by averaging the negative gradients. A further set of features relate to amplitude; \textit{Min Amp}, \textit{Avg Amp} and \textit{Max Amp}. Amplitude is the increase in value that is measured between an up peak and its previous point, divided by the \textit{Max Y} value. That means, it is the increase relative to the maximum value. An example is shown at around week 250 in (b). The amplitude is labelled in the middle of the plot and for the purpose of the example, the difference between the peak and the previous value is equal to 1000. The \textit{Max Y} value of the time-series is roughly 4000, which therefore means an amplitude of 25\%. \textit{Min Amp} is the lowest measured amplitude, \textit{Avg Amp} is the mean of all amplitudes, and \textit{Max Amp} is the highest amplitude.

 A complete list of the 42 extracted features is presented in Table \ref{tab:features}. The measures are extracted separately, which means that there is one feature vector per measure. In Table \ref{table:git_log_features_example}, a sample of the extracted features is presented. Some features cannot be extracted from all measures. For example, peak-related features cannot be measured on time-series with a length of three or less, because peaks cannot be detected. Where features are immeasurable the value is set to zero (“0”) during the preprocessing step (Step E) because the k-means algorithm cannot handle missing values. Each repository in the input collection is processed in this way. After this, feature vectors are used in the subsequent steps. 
 
Some of the features capture characteristics that might depend on the project size (e.g. Sum y, Max. y). However, due to the manual investigation done by the baseline study, we guess that larger projects better meet the definition of engineered project and would rarely aim to deliver toy applications or solutions to students' homework assignments. Also, there are many features extracted by PHANTOM that should not depend on the size of the projects but rather on the changes in the ``intensity'' of product development (e.g. Min. TBP up, Max. TBP up).

\newcommand{\lenCategory}{1.8cm}
\newcommand{\lenDescription}{7cm}
\begin{table}
    \centering
    \caption[PHANTOM Features]{Features extracted by PHANTOM}
    \label{tab:features}
    \begin{tabular}{p{\lenCategory} | l p{\lenDescription}}
        \hline\noalign{\smallskip}
    
    Duration                    & Duration & Time interval in weeks between the first and last week of the time-series\\
    \noalign{\smallskip}\hline\noalign{\smallskip}
    \multirow{8}{\lenCategory}{Y value}    & Max. y & The highest value\\
                                & Max. y Pos & The week number of the Max Y\\
                                & Mean y & The average value\\
                                & Sum y & The sum of all values\\
                                & q25 & The 25\% quantile of values\\
                                & q50 & The 50\% quantile of values\\
                                & q75 & The 75\% quantile of values\\
                                & std & The standard deviation of values\\
    \noalign{\smallskip}\hline\noalign{\smallskip}
    \multirow{3}{\lenCategory}{Peaks}        & Peak down & The number of downwards facing peaks\\
                                & Peak none & The number of peaks that are neither downwards, nor upwards facing peaks\\
                                & Peak up   & The number of upwards facing peaks\\
    \noalign{\smallskip}\hline\noalign{\smallskip}
    \multirow{6}{\lenCategory}{Time between peaks} & Min. TBP up & \multirow{6}{\lenDescription}{The time between upwards facing peaks is measured as the number of weeks between two neighbouring peaks.}\\
                                        & Avg. TBP up & \\
                                        & Max. TBP up & \\
                                     &  Min. TBP down & \\
                                        & Avg. TBP down & \\
                                        & Max. TBP down & \\
    \noalign{\smallskip}\hline\noalign{\smallskip}
    \multirow{3}{\lenCategory}{Amplitude}     & Min. amplitude & \multirow{3}{\lenDescription}{The amplitude is the difference in height between a peak and the previous valley. This value is normalised by dividing it with Max Y.}\\
                                & Avg. amplitude & \\
                                & Max. amplitude & \\
                                
    \noalign{\smallskip}\hline\noalign{\smallskip}
    \multirow{6}{\lenCategory}{Positive and negative peak deviation}     & Min. PPD & \multirow{3}{\lenDescription}{The positive peak deviation (PPD) is the difference between the Mean Y value and the y value of a upwards facing peak.}\\
                                                            & Avg. PPD & \\
                                                            & Max. PPD & \\
                                                            & Min. NPD & \multirow{3}{\lenDescription}{The negative peak deviation (NPD) is the difference between the Mean Y value the y value of a downwards facing peak.}\\
                                                            & Avg. NPD & \\
                                                            & Max. NPD & \\
                                                            
    \noalign{\smallskip}\hline\noalign{\smallskip}
    \multirow{8}{\lenCategory}{Positive and negative sequences}& Min. PS & \multirow{8}{\lenDescription}{A sequence is, when at least two sequential gradients have the same sign. Therefore, the positive (PS) and negative sequences (NS) are numeric values, that count the number of sequential same sign gradients.}\\
                                                    & Avg. PS & \\
                                                    & Max. PS & \\
                                                    & Sum. PS & \\
                                                    & Min. NS & \\
                                                    & Avg. NS & \\
                                                    & Max. NS & \\
                                                    & Sum. NS & \\
                                                    
    \noalign{\smallskip}\hline\noalign{\smallskip}
    \multirow{8}{\lenCategory}{Positive and negative gradients}& Min. PG & \multirow{6}{\lenDescription}{Gradients are the difference between two neighbouring y values.}\\
                                                    & Avg. PG & \\
                                                    & Max. PG & \\
                                                    & Min. NG & \\
                                                    & Avg. NG & \\
                                                    & Max. NG & \\ \noalign{\smallskip}\cline{2-3}\noalign{\smallskip}
                                                    & PG Count & Number of positive gradients (PG)\\
                                                    & NG Count & Number of negative gradients (NG)\\
    \noalign{\smallskip}\hline
    \end{tabular}
\end{table}
 
\begin{table}
    \centering
    \caption[Git log as a features example]{Example feature vectors for the five measures.}
    \label{table:git_log_features_example}
    \begin{tabular}{l | ccc}
    \hline\noalign{\smallskip}
    Measure & Duration & Avg Y & Max Y \\
    \noalign{\smallskip}\hline\noalign{\smallskip}
    Integrations                & 2                            & 1.5                       & 2                         \\ [1ex]
    Integrators                 & 2                            & 1.5                       & 2                         \\ [1ex]
    Commits                     & 2                            & 1.5                       & 2                         \\ [1ex]
    Authors                     & 2                            & 1                         & 1                         \\ [1ex]
    Merges                      & 2                            & 0                         & 0 \\
    \noalign{\smallskip}\hline
    \end{tabular}
    
\end{table}

The next step is to select the best subset of features from the feature vector (Step F). In order to remove redundant features, the Pearson correlation coefficient is calculated. If the correlation meets or exceeds a specified threshold the feature is removed.

The remaining features are normalised with the standard score and then used to fit a k-means model (Step G). In this study, we divided observations into two clusters and used the standard configuration of k-means in the Python library Scikit-Learn:\footnote{\cite{scikit-learn}, \url{https://scikit-learn.org}}
\begin{itemize}
\item number of clusters = 2,
\item number of initialisations = 10,
\item centroid update algorithm = k-means++ (Lloyd's algorithm),
\item max iterations = 300.
\end{itemize}

Finally, the fitted model is outputted. A new observation is classified to a cluster by measuring its Euclidean distance to all centroids and assigning it to the cluster represented by the closest one. 

Although PHANTOM uses unsupervised models, it is worth to emphasise that the features extracted and preprocessed in steps D and E could be used as an input to supervised models, if the ground truth is available for the considered dataset.

\section{Validation Results}
\label{sec:validation}

We validated PHANTOM against each of the requirements defined in Section \ref{sec:problem_investigation}.  For the requirements requiring evaluating the accuracy of the proposed method or comparing it with the baseline study (\emph{reaper}), we based the validation on the datasets and prediction quality measures presented in Section \ref{sec:methodology}.

\subsection{All Measures Must Be Extractable From The Git Log (Req1)}
\label{sec:validation_R1}

The five measures extracted by PHANTOM are Integration Frequency, Commit Frequency, Integrator Frequency, Committer Frequency, and Merge Frequency. Each measure uses different information from the Git log, along with at least one of the two date types (author or committer date). The other parts of the Git log are the author and committer email, and the number of parent commits (see Table \ref{table:phantom_measures}). All of this information is available in every Git managed repository, and, in fact, Git ensures its availability, because when committing changes,  the information is automatically recorded. PHANTOM has no dependency on additional data from GitHub, such as the GitHub API and mirroring services like GHTorrent. Due to the decision to use this specific set of information, PHANTOM is able to extract all measures from Git logs exclusively.

\subsection{Time-Series Must Be Characterised By Feature Vectors Accurately (Req2)} 
\label{sec:validation_R2}

\begin{figure}
    \centering
\includegraphics[width=\linewidth,keepaspectratio]{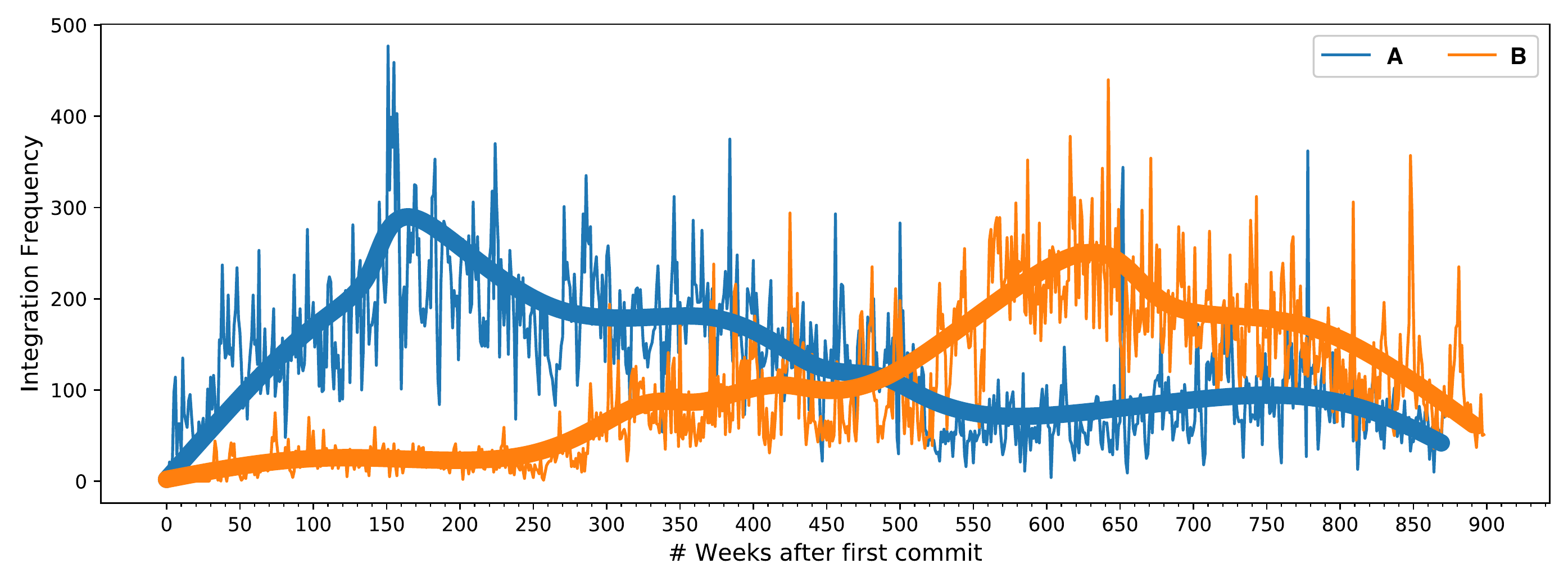}
    \caption[Two Integration Frequencies]{Integration Frequencies for \url{https://github.com/mono/mono} (A) and  \url{https://github.com/FFmpeg/FFmpeg} (B).}
    \label{fig:patterns1_overlay}
\end{figure}

Feature vectors must capture the characteristics of time-series accurately to allow k-means to find meaningful clusters. It can be difficult to select features that do this. This problem is illustrated by the two integration frequencies plotted in Figure \ref{fig:patterns1_overlay}, which were selected from an investigation of twenty repositories using PHANTOM. The plots are visually distinguishable. However, when converted into feature vectors, a small number of features, such as \textit{Max Y} or \textit{Duration} may not be enough to differentiate two time-series from each other (see Table \ref{table:timeseries_features_example}). The difference between the \textit{Max Y} values and the \textit{Duration} values is 30 and 40 respectively. \textit{Max Y} and \textit{Duration} are close enough that one can say that the time-series are similar to each other. Therefore, crucial features are missing to differentiate them. An additional feature such as the x value of the highest peak (\textit{Max Y Pos}) would show a clear difference between the two repositories. PHANTOM uses \textit{Duration}, \textit{Max Y}, and \textit{Max Y Pos} along with 40 additional features described in Table \ref{tab:features} to characterise time-series.

\begin{table}
    \centering
    \caption[Example Feature Vectors for Integration Frequencies]{Example feature vector for the time-series in Figure \ref{fig:patterns1_overlay} (the values have been rounded).}
    \label{table:timeseries_features_example}
    \begin{tabular}{c | ccc}
	\hline\noalign{\smallskip}
        Repository & Duration & Max Y & Max Y Pos \\ 
\noalign{\smallskip}\hline\noalign{\smallskip}
        A (\url{https://github.com/mono/mono})       & 870      & 470   & 160            \\
        B (\url{https://github.com/FFmpeg/FFmpeg})          & 900      & 430   & 640           \\
\noalign{\smallskip}\hline
    \end{tabular}%

\end{table}

Even if a number of features are similar, the chances of an identical feature vector for two different time-series is reduced with a larger feature vector. By this, the k-means algorithm is able to cluster time-series via feature vectors effectively, as those differences are clear. By using larger feature vectors, PHANTOM captures the characteristics of time-series and the differences between them are highlighted.

\subsection{The Established Ground Truth Can Be Discovered Using Unsupervised Learning (Req3)}
\label{sec:validation_R3}

We used PHANTOM to fit k-means models on the ground-truth datasets\footnote{The features extracted by PHANTOM from the datasets and supplemental materials are published under DOI: 10.5281/zenodo.3483755.}. These are the Organization and Utility datasets, which are both complemented with the negative instances so they contain engineered and not engineered repositories to almost equal parts. PHANTOM requires a correlation threshold to select the best subset of features. As it is unknown which threshold is the best, we explored thresholds ranging from 0.05 to 1, with a step size of 0.05. This means that for each combination of datasets and measures, twenty models were fitted. As k-means is unsupervised, the true labels are not known to the algorithm when fitting the model, which enables a comparison of the produced cluster labels and the ground-truth labels. 

\begin{figure}
    \includegraphics[width=\linewidth, keepaspectratio]{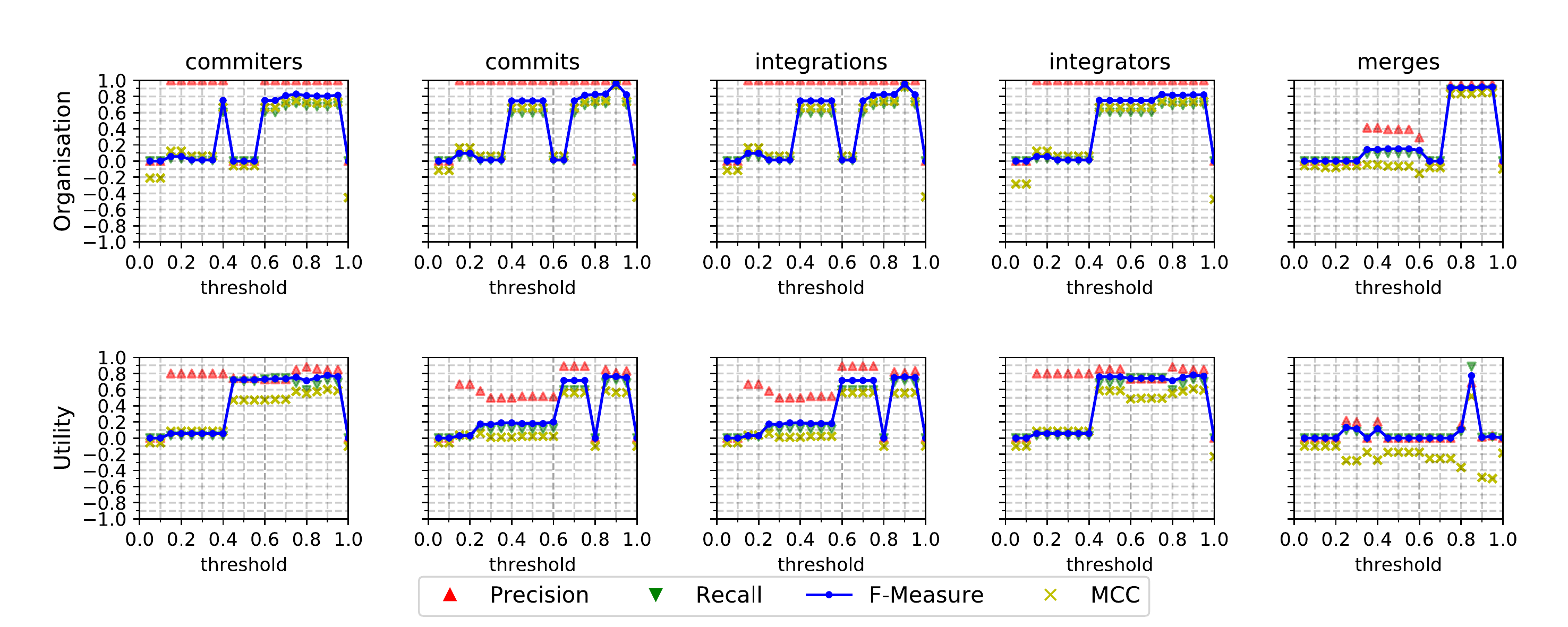}
    \caption[Accuracy on Ground Truth]{The accuracy of k-means model thresholds against the the \emph{ground truth} for the five measures (the Organization and Utility datasets combined with the Negative Instances dataset).}
    \label{fig:training_accuracy}
\end{figure}

The comparisons of obtained prediction quality measures of Organization and Utility datasets depending on the correlation threshold are presented in Figure \ref{fig:training_accuracy}.  On the Organization dataset, there are many models that achieve Precision and Recall close to 1.0. On the utility dataset, the accuracy is lower with Precision and Recall of up to 0.9. The high Precision and Recall indicate that the models were able to rediscover the majority of true labels for both datasets. Overall, the Organization repositories could be rediscovered with higher accuracy than the Utility repositories. However, the accuracy largely depends on the dataset, measure and correlation threshold. This shows that a ground truth could successfully be discovered using an alternative, unsupervised technique.

Based on the results presented in Figure \ref{fig:training_accuracy}, we can also observe that setting a threshold anywhere between 0.75 and 0.95 will result in obtaining very similar accuracy of all the models. 

\subsection{The method Performs Well On Commodity Hardware At Large-Scale (Req4)}
\label{sec:validation_R4}

In order to validate the applicability of PHANTOM to process large-scale data, we performed two analyses. In the first one, we used PHANTOM to download and process the labelled, smaller datasets; Utility, Negative Instances, and Validation to extrapolate the results to the size of the Large dataset. Since the Organization dataset comes from a different selection process than the other datasets, it is not used for this extrapolation.  The second analysis is performed directly on the Large dataset.

PHANTOM's results for both analyses are presented in Table \ref{table:timings}. For the smaller datasets, it shows that 3.8\% of the repositories were unavailable, which means they have been deleted or made private. The total download time was 7.5 minutes for 500 repositories, which gives the average repository-download time of 0.94 seconds. When these values are extrapolated up to the Large dataset (1,857,423 software project repositories), the time to obtain the Git logs is estimated to be 20.2 days. 

\begin{table}
    \centering
        \caption[PHANTOM Timings]{Number of available repositories and timings to clone and generate the Git log for them.}
    \label{table:timings}
    \begin{tabular}{l|cc}
    \hline\noalign{\smallskip}
    Dataset & Available Repositories & Time Taken  \\ 
    \noalign{\smallskip}\hline\noalign{\smallskip}
    Organization                    & 149 / 150              & 10:39 minutes    \\
    Utility                         & 145 / 150              & 2:44 minutes       \\
    Negative Instances              & 138 / 150              & 1:53 minutes       \\
    Validation(engineered)     & 100 / 100              & 1:36  minutes      \\
    Validation(Not engineered) & 98 / 100               & 1:18  minutes      \\
Large & 1,780,773 / 1,857,423 &  21.5 days \\
    \noalign{\smallskip}\hline
    \end{tabular}%
\end{table}

For the Large dataset, it took PHANTOM 21.5 days to obtain the Git logs for 1,780,773 (95.36\% of all) repositories. This leaves 76,650 (4.64\% of all) repositories that were not available, due to either deletion or being made private. When converting the Git logs to time-series, some of the logs had to be excluded from the analysis, because of a formatting problem; Logs with author and committer names that contain a comma (“,”) had to be excluded, because the additional comma made the correct separation of information impossible since Git logs are saved as CSV files. For this reason 9,606 (0.5\% of the obtained) Git logs had to be discarded. The remaining 1,771,167 Git logs were converted to time-series and feature vectors were extracted.

\subsection{The Method Provides Comparable Accuracy To Supervised Methods (Req5)}
\label{sec:validation_R5}

\begin{figure}
    \includegraphics[width=\linewidth, keepaspectratio]{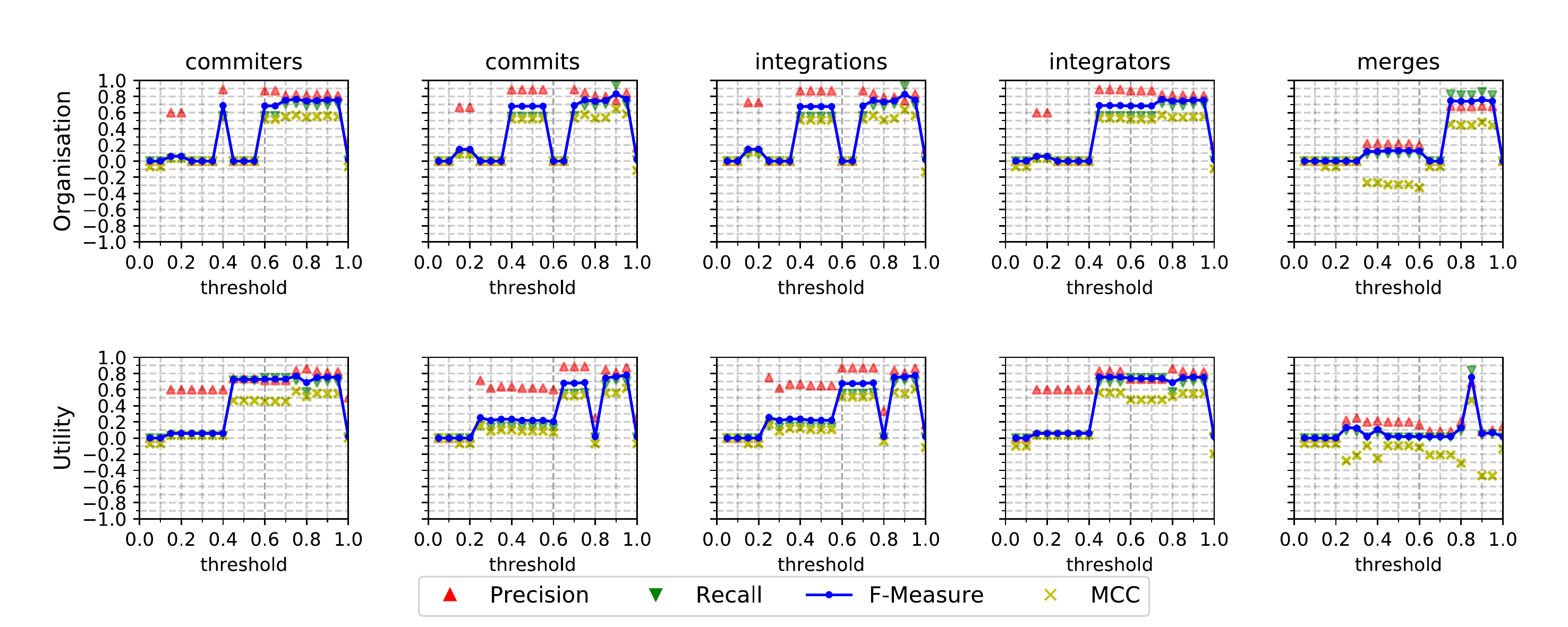}
    \caption[Accuracy on Validation Data]{The accuracy of k-means model thresholds against the the \emph{Validation} data for the five measures (fitted on the Organization and Utility datasets combined with the Negative Instances dataset).}
    \label{fig:validation_accuracy}
\end{figure}

In the baseline study, custom Score-based and Random Forest classifiers were trained on the Organization and Utility ground-truth datasets (each combined with the instances from the Negative Instances dataset). The classifiers were then used to predict the Validation dataset. In order to compare k-means to these algorithms, PHANTOM is applied to the same datasets. First, similarly to Section \ref{sec:validation_R3}, PHANTOM explores a range of thresholds for each combination of datasets and measures. This time, however, the fitted models are used to predict repositories from the Validation dataset. Accuracy when predicting repositories is shown in Figure \ref{fig:validation_accuracy}. As already seen in Figure \ref{fig:training_accuracy}, the accuracy varies across measures, datasets, and thresholds. Although we base our discussion of the PHANTOM accuracy based on the results obtained for the best-performing PHANTOM models, we can see in Figure \ref{fig:validation_accuracy} that choosing any threshold between 0.75 and 0.95 would result in models having very similar filtering accuracy. 

To compare the results against the baseline study, the best models for each dataset and measure are selected. In order to achieve this, the authors established a set of rules that determine the best model:
\begin{enumerate}
\item Find the highest F-Measure.
\item Find the highest Precision.
\item Find the highest Recall.
\item Find the lowest correlation threshold.
\end{enumerate}

These rules are implemented in PHANTOM which automates the feature selection process. The best models are presented in tables \ref{table:accuracy_best_organisation} and \ref{table:accuracy_best_utility}.

\begin{table}
    \centering
    \caption[Prediction Accuracy (Organization)]{Prediction accuracy for the five measures when clustering repositories. (Organization models)}
    \label{table:accuracy_best_organisation}
    \begin{tabular}{lc | ccc | c | c}
    \hline\noalign{\smallskip}
        Measure      & Threshold & Precision & Recall & F-Measure & MCC  & Features \\ 
        \noalign{\smallskip}\hline\noalign{\smallskip}
        Merges       &   0.90    & 0.68      &  0.85  &   0.76    & 0.48 & 17       \\
        Integrators  &   0.75    & 0.82      &  0.71  &   0.77    & 0.57 & 9        \\
        Integrations &   0.90    & 0.75      &  0.93  &   0.83    & 0.64 & 19       \\
        Commits      &   0.90    & 0.75      &  0.94  &   0.83    & 0.65 & 19       \\
        Committers   &   0.75    & 0.82      &  0.71  &   0.77    & 0.57 & 11\\
        \noalign{\smallskip}\hline
    \end{tabular}
    
\end{table}

\begin{table}
    \centering
    \caption[Prediction Accuracy (Utility)]{Prediction accuracy for the five measures when clustering repositories. (Utility models)}
    \label{table:accuracy_best_utility}
    \begin{tabular}{lc | ccc | c | c}
    \hline\noalign{\smallskip}
        Measure      & Threshold & Precision & Recall & F-Measure & MCC  & Features \\ 
        \noalign{\smallskip}\hline\noalign{\smallskip}
        Merges       &   0.85    & 0.68      &  0.84  &   0.75    & 0.47 & 11       \\
        Integrators  &   0.45    & 0.83      &  0.69  &   0.76    & 0.56 & 5        \\
        Integrations &   0.95    & 0.86      &  0.70  &   0.78    & 0.61 & 22       \\
        Commits      &   0.95    & 0.87      &  0.70  &   0.78    & 0.62 & 22       \\
        Committers   &   0.75    & 0.83      &  0.71  &   0.77    & 0.58 & 8\\
        \noalign{\smallskip}\hline
    \end{tabular}
    
\end{table}

Before comparing the accuracy of PHANTOM models with the classifiers from the baseline study, we first compare their accuracy to the accuracy of three na\"ive classifiers (uniform random, stratified random, and selecting the most frequently appearing class). The F-Measure for these classifiers were equal to 0.52, 0.51, and 0.67, respectively. By looking at the results presented in tables \ref{table:accuracy_best_organisation} and \ref{table:accuracy_best_utility}, we can see that all of the best-performing PHANTOM models visibly outperformed the na\"ive classifiers. 

The accuracy of the models proposed in the baseline study is presented in Table \ref{table:accuracy_munaiah}. The baseline classifiers, trained on the Organization dataset, achieves a lower F-Measure than any PHANTOM model fitted to the same data. PHANTOM matches the Precision and surpasses the Recall of the supervised algorithms. Classifiers trained on the Utility dataset set a higher benchmark than classifiers trained on the Organization dataset. PHANTOM matches the F-Measure of the score-based classifier. Although the highest Precision on the Utility dataset of 0.82 could be surpassed by two out of five PHANTOM models (Integrations, Commits), the Recall on these was lower than the one obtained for the RF classifier. Interestingly, the RF classifier trained on the Organization dataset in the baseline study performed visibly worse than one of the na\"ive classifiers. Also, the Score-based classifier trained on the same dataset achieved an F-Measure only 0.01 higher than the na\"ive classifier. 

\begin{table}
    \centering
    \caption[Prediction Accuracy (Baseline)]{The accuracy of the baseline's tool \textit{reaper} as reported in \cite{Munaiah2017}.}
    \label{table:accuracy_munaiah}    
    \begin{tabularx}{0.8\linewidth}{l l | c c c}    
    \hline\noalign{\smallskip}
        Data set     & Classifier    & Precision    & Recall    & F-Measure    \\ 
        \noalign{\smallskip}\hline\noalign{\smallskip}
        Organization& Score-based    & 0.76        & 0.61        & 0.68        \\
        Organization& Random Forest & 0.88        & 0.42        & 0.57        \\ 
        \noalign{\smallskip}\hline\noalign{\smallskip}
        Utility        & Score-based    & 0.58        & 0.99        & 0.73        \\
        Utility        & Random Forest    & 0.82        & 0.86        & 0.84        \\
        \noalign{\smallskip}\hline
    \end{tabularx}
    
\end{table}

As a follow-up analysis, we trained a set of Random Forest classifiers\footnote{We used the default configuration of Random Forest provided by the Scikit-Learn library since \cite{Munaiah2017} did not provide information about the RF meta-parameters they used in their study.} using the measures and extracted features provided by PHANTOM to investigate whether the features allow obtaining similar results to the baseline study.  The prediction accuracy of the RF Classifiers are presented in tables \ref{table:rf_accuracy_best_organisation} and \ref{table:rf_accuracy_best_utility}.  The F-Measure scores obtained for the best RF models for the Organization and Utility datasets were equal to 0.72 and 0.79 respectively.  For the Organization dataset, the F-Measure was higher by 0.15 when compared to the corresponding RF models in the baseline study, while for the second dataset, it was lower by 0.05.  This confirms that the features extracted by PHANTOM allow separating instances of engineered and not engineered projects for both unsupervised and supervised models.  Nevertheless, when compared to the accuracy of  PHANTOM, the F-Measure scores for PHANTOM were higher by 0.11 for the first dataset and lower by 0.01 for the second one than for the RF classifiers trained using the same features. Thus, it seems that the unsupervised model used by PHANTOM can achieve higher prediction accuracy than RF for some cases.

\begin{table}
    \centering
    \caption[Prediction Accuracy (Organization)]{Prediction accuracy for the five measures and Random Forest classifiers. (Organization models)}
    \label{table:rf_accuracy_best_organisation}
    \begin{tabular}{l | ccc | c}
    \hline\noalign{\smallskip}
        Measure      & Precision & Recall & F-Measure & MCC  \\ 
        \noalign{\smallskip}\hline\noalign{\smallskip}
        Merges       & 0.95      &  0.58  &   0.72    & 0.59       \\
        Integrators  & 0.94      &  0.44  &   0.60    & 0.48         \\
        Integrations & 0.94      &  0.47  &   0.63    & 0.51        \\
        Commits      & 0.93      &  0.38  &   0.54    & 0.43        \\
        Committers   & 0.93      &  0.42  &   0.58    & 0.46 \\
        \noalign{\smallskip}\hline
    \end{tabular}
    
\end{table}

\begin{table}
    \centering
    \caption[Prediction Accuracy (Utility)]{Prediction accuracy for the five measures and Random Forest classifiers. (Utility models)}
    \label{table:rf_accuracy_best_utility}
    \begin{tabular}{l | ccc | c}
    \hline\noalign{\smallskip}
        Measure      & Precision & Recall & F-Measure & MCC \\ 
        \noalign{\smallskip}\hline\noalign{\smallskip}
        Merges       & 0.74      &  0.78  &   0.76    & 0.51        \\
        Integrators  & 0.78      &  0.75  &   0.77    & 0.54         \\
        Integrations & 0.78      &  0.80  &   0.79    & 0.57        \\
        Commits      & 0.77      &  0.79  &   0.78    & 0.56        \\
        Committers   & 0.80      &  0.75  &   0.77    & 0.56 \\
        \noalign{\smallskip}\hline
    \end{tabular}
    
\end{table}

Finally, we compared the accuracy of the models on the sample of the Large dataset (see Section \ref{sec:val-eval-methodology}). The baseline study concluded that the number of engineered projects in the Large dataset is 24\%, which was the prediction made by the best-performing model (RF trained on the Utility dataset). The predictions made by other models differed visibly and ranged from 6\% to 70\%. We estimated the accuracy of the best model to be ca. 0.72 with the 0.95 Confidence Interval (CI)\footnote{We first calculated Accuracy for each stratum, and then aggregated them using the weighted average with weights being the numbers of observations belonging to strata in the Large dataset.} between 0.57 and 0.83.

We selected the PHANTOM models that performed best on the Validation dataset and applied them to classify the projects in the Large dataset. This means that ten models (one for each combination of ground-truth dataset and measure) were used to predict the repository labels for the Large dataset. In tables \ref{table:prediction_organisation} and \ref{table:prediction_utility}, the best models and the number of repositories predicted to be engineered are presented. Out of these models, six resulted in a percentage of engineered repositories between 35\% and 40\% and two resulted in 55\%. The remaining two models resulted in 19\% and 96\%. Accuracy for all the PHANTOM models but one was slightly lower than the best model from the baseline study (ca. -0.03).

\begin{figure}
    \includegraphics[width=\linewidth, keepaspectratio]{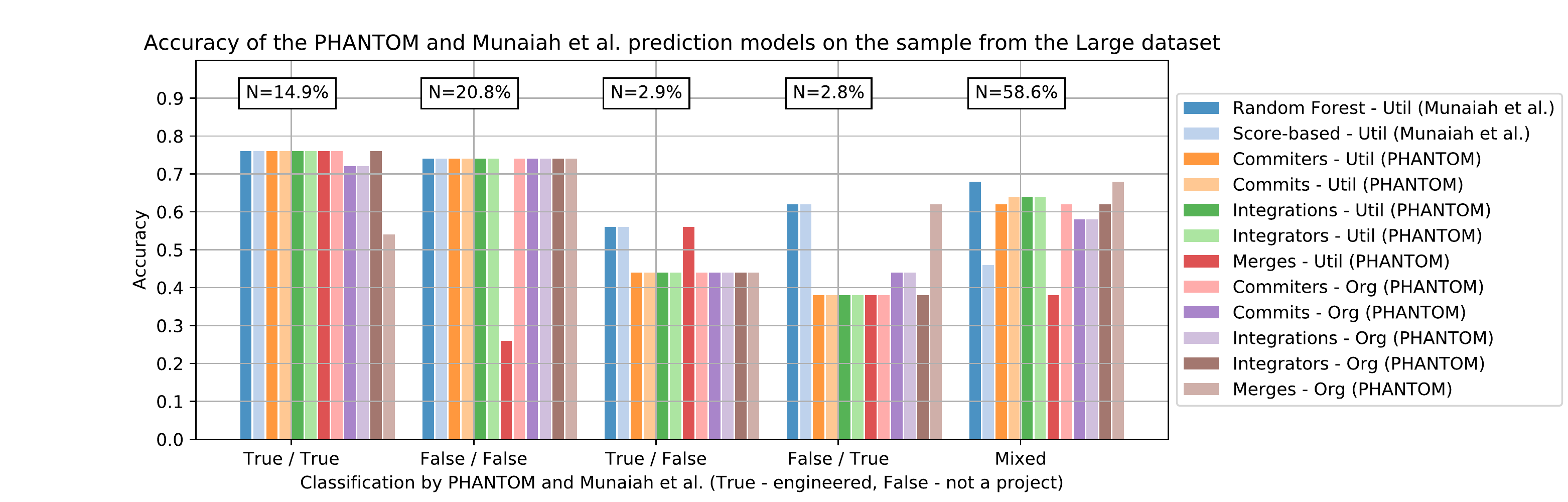}
    \caption{Accuracy for the baseline and PHANTOM models on the sample from the Large dataset.}
    \label{fig:acc-large}
\end{figure}

Figure \ref{fig:acc-large} shows the comparison between the accuracy of the baseline and PHANTOM models for a sample of the repositories from the Large dataset. We can see that whenever the models were unanimous in predictions (for ca. 35\% of the repositories) they achieved the highest accuracy ca. 0.75. On the contrary, when the baseline and PHANTOM models strongly disagreed, their accuracy dropped visibly, even to the level of random guessing. However, there should be only around 6\% of such instances. For the remaining cases, the accuracy of the best models ranged between 0.6 and 0.7 (for ca. 59\% of the repositories).

The most accurate PHANTOM model predicted the number of engineered repositories to be 19\%, which is quite similar to the prediction made by the best model from the baseline study. However, it is important to emphasise that the predictions made in both studies could be affected by a visible error because the accuracy of the models was ca. 0.70. 

Most of the false-positive repositories contained student assignments, automatically exported code, private sandboxes, and web applications with no documentation (usually developed using the Ruby on Rails and Django frameworks).  Many of these repositories contained code of applications or services, however, they did not provide enough information to decide whether the application is intended to be used by anyone else than the authors. Many of the False Positives are repositories for student assignment projects. They are often well-structured and documented. However, the main question is whether the outcomes of such projects may attract external users. When it comes to False Negatives, they were dominated by the repositories containing very small utility tools (e.g. often having the form of a single script), showcase applications, and plugins/extensions.

\begin{table}
	\centering
		\caption[PHANTOM Results Organization]{Results from PHANTOM on 1,771,167 repositories from the Large dataset. Each row represents one model (Organization models).}
	\label{table:prediction_organisation}
	\begin{tabular}{l c | c c c c}
	\hline\noalign{\smallskip}
	Measure      & 
	Threshold & 
	\# of engineered & 
	\% of engineered & Acc & Acc 0.95 CI\\ 
	\noalign{\smallskip}\hline\noalign{\smallskip}
	
	Merges       & 0.9       & 343,818                     & 19\%  & 0.69  & (0.53, 0.79)                      \\
	Commits      & 0.9       & 704,995                     & 40\% & 0.65    & (0.49, 0.76) \\
	Committers   & 0.75      & 688,260                     & 39\% & 0.67   & (0.52, 0.78)                     \\
	Integrations & 0.9       & 700,501                     & 39\% & 0.65   & (0.49, 0.76),                     \\
	Integrators  & 0.75      & 688,260                     & 38\%  & 0.67  & (0.52, 0.78) \\
	\noalign{\smallskip}\hline
	\end{tabular}%

\end{table}

\begin{table}
	\centering
		\caption[PHANTOM Results Utility]{Results from PHANTOM on 1,771,167 repositories from the Large dataset. Each row represents one model (Utility models).}
	\label{table:prediction_utility}
	\begin{tabular}{l c | c c c c}
	\hline\noalign{\smallskip}
	Measure      & 
	Threshold & 
	\# of engineered & 
	\% of engineered & Acc & Acc 0.95 CI\\ 
	\noalign{\smallskip}\hline\noalign{\smallskip}
	Merges       & 0.85      & 1,716,118                   & 96\%   & 0.41   &   (0.29, 0.55)                 \\
	Commits      & 0.95      & 985,213                     & 55\%  & 0.68    & (0.54, 0.79)                   \\
	Committers   & 0.75      & 690,371                     & 39\%  & 0.67   &  (0.52, 0.78)                   \\
	Integrations & 0.95      & 983,553                     & 55\%  & 0.68   &  (0.54, 0.79)                   \\
	Integrators  & 0.45      & 617,042                     & 35\%  & 0.68 & (0.54, 0.79) \\  
	\noalign{\smallskip}\hline
	\end{tabular}%

\end{table}

\subsection{The Method Can Be Used To Filter Projects In Different Contexts  (Req6)}
\label{sec:validation_R6}

When the best-performing classifiers for each of the measures were applied to the Industry dataset, they predict that the percentage of repositories containing engineered software projects is between 12\% and 95\%. The most accurate PHANTOM model (Merges, Organization) classified repositories with F-Measure and MCC equal to 0.89 and 0.53, respectively.

We discussed the results with the companies' employees. They categorised the Git repositories based on their contents. The resulting categories are presented in Table \ref{tab:pred-industry}.  It turned out that only 35 of the repositories contained source code of products developed for external customers. According to the employees, these repositories should be identified as containing engineered projects. As it is shown in Table \ref{tab:pred-industry}, most of the classifiers correctly indicated that the majority of the repositories belong to this decision class (the average accuracy was equal to 0.70).  The best-performing model on the Large dataset (Merges, Organization) correctly recognised 94\% of them as containing engineered projects while the corresponding model trained on the Utility dataset (Merges, Utility) was able to correctly classify all of them. 

The second group of categories contains repositories storing internal libraries and reusable components which are parts of products developed by Company A (UI component, Ruby gems or extensions, npm packages, or Python libraries).  As we can see, most of these repositories were correctly indicated as the ones containing engineered projects. Repositories that were misclassified as not engineered software projects were mostly micro-projects having short or irregular development histories. 

The third category contains repositories storing examples, snippets, spike solutions, and template solutions. According to the employees, nearly all of these repositories should be classified as containing not engineered projects. PHANTOM models correctly recognized ``toy projects'', however, they misclassified not-engineered projects that had long and active histories of development (e.g. a pattern application).

The next category groups forked repositories of external libraries that were modified and used by Company A. These repositories were correctly indicated as containing engineered projects. However, it shows that there is a need for manually identifying the forks of repositories to avoid including clones of repositories when conducting MSR studies on the GitHub data since they can bias the results. 

The fifth group of categories includes repositories that store source code of applications developed during Hackathon sessions or being the result of internal Research \& Development. These were mostly considered by the employees as not engineered projects and correctly identified as such by the PHANTOM models.  However, there were a few projects that have matured over time and are now regularly used in Company A (categorised as internal APIs and internal applications). The best-performing PHANTOM model on the Large dataset correctly classified 1 out of 2 internal API projects and 83\% of internal applications. 

The sixth group of categories contains repositories storing the code of internal tools (scripts), infrastructure code (e.g. Docker), or setup (configurations). Although some of the tools could be considered as engineered software projects, most of the infrastructure and setup repositories should be regarded as not engineered. All of the projects belonging to these categories were correctly classified by the best-performing PHANTOM classifier.

Finally, the remaining category groups repositories containing documentation (usually in the form of markdown files). Interestingly, some of the repositories were falsely identified as containing engineered projects. Those that were most frequently misclassified contained knowledge bases that are regularly used and updated within Company A, making their commit histories similar to active software development projects. This revealed a limitation of relying on Git commit frequencies when training classifiers instead of analysing multiple project artefacts. Therefore, it might be worth to extend the list of features to include the extensions of the files being modified in the project to filter such cases.

\begin{table}[!ht]

\setlength{\tabcolsep}{2pt}
\renewcommand{\arraystretch}{1.3}
    \centering
    \caption{The predictions made by the PHANTOM models for the Industry dataset}
    \label{tab:pred-industry}    
    
    \begin{adjustbox}{angle=90}
    \begin{tabular}{c | r | c | c | c | c c c c c | c c c c c }    
    \hline\noalign{\smallskip}
    \multicolumn{2}{c |}{ } & &  & & \multicolumn{5}{c |}{Organization dataset} & \multicolumn{5}{c }{Utility dataset} \\
    
    \multicolumn{2}{c |}{} & & \# of &  &  & Inte- & Inte- &  & Commi- &  & Inte- & Inte- &  & Commi- \\
        
        \multicolumn{2}{c |}{Type} & N &  eng. & Average & Merges & grators & grations & Commits & ters & Merges & grators & grations & Commits & ters \\
        
\noalign{\smallskip}\hline\noalign{\smallskip} 
 \multicolumn{2}{r |}{F-Measure} & \multirow{4}{*}{100} & \multirow{4}{*}{75} &  0.73      & 0.89 & 0.81 & 0.84 & 0.84 & 0.81 & 0.85 & 0.18 & 0.84 & 0.84 & 0.36\\ 
\multicolumn{2}{r |}{ Recall} &   & & 0.72      & 0.91 & 0.79 & 0.83 & 0.84 & 0.77 & 0.96 & 0.11 & 0.88 & 0.89 & 0.24 \\
\multicolumn{2}{r |}{Precision} &   & & 0.80      & 0.87 & 0.84 & 0.85 & 0.84 & 0.84 & 0.76 & 0.67 & 0.80 & 0.80 & 0.75 \\
\multicolumn{2}{r |}{MCC} &   & & 0.24  & 0.53 & 0.33 & 0.38 & 0.36 & 0.31 & 0.08 & -0.07 & 0.23 & 0.25 & 0.00 \\
\noalign{\smallskip}\hline\noalign{\smallskip}
\parbox[t]{5mm}{\multirow{18}{*}{\rotatebox[origin=c]{90}{Accuracy}}}
& Project & 35 & 35 &  0.70      & 0.94 & 0.66 & 0.86 & 0.89 & 0.60 & 1.00 & 0.03 & 0.89 & 0.91 & 0.23\\\cline{2-15}
& UI Component & 4 & 4 &  0.70      & 0.75 & 1.00 & 0.50 & 0.50 & 1.00 & 1.00 & 0.50 & 0.75 & 0.75 & 0.25 \\
& Ruby gem & 5 & 4 &  0.60      & 0.80 & 0.80 & 0.80 & 0.80 & 0.80 & 0.80 & 0.00 & 0.60 & 0.60 & 0.00 \\
& npm package & 1 & 1 &  0.80      & 1.00 & 1.00 & 1.00 & 1.00 & 1.00 & 1.00 & 0.00 & 1.00 & 1.00 & 0.00 \\
& Extension & 3 & 2 &  0.73      & 0.67 & 1.00 & 0.67 & 0.67 & 1.00 & 0.67 & 0.67 & 0.67 & 0.67 & 0.67 \\
& Library & 1 & 1 &  0.80      & 1.00 & 1.00 & 1.00 & 1.00 & 1.00 & 1.00 & 0.00 & 1.00 & 1.00 & 0.00\\\cline{2-15}
& Example/Snippet & 9 & 2 &  0.44      & 0.33 & 0.33 & 0.33 & 0.33 & 0.44 & 0.44 & 0.78 & 0.33 & 0.33 & 0.78 \\
& Pattern app & 1 & 0 &  0.20      & 0.00 & 0.00 & 1.00 & 0.00 & 0.00 & 0.00 & 1.00 & 0.00 & 0.00 & 0.00 \\
& Spike solution & 2 & 0 &  0.55      & 0.50 & 0.50 & 0.50 & 0.50 & 0.50 & 0.00 & 1.00 & 0.50 & 0.50 & 1.00\\\cline{2-15}
& Fork & 12 & 12 &  0.82      & 1.00 & 0.92 & 1.00 & 1.00 & 1.00 & 0.75 & 0.08 & 0.92 & 0.92 & 0.58\\\cline{2-15}
& Hackathon & 2 & 0 &  0.90      & 1.00 & 1.00 & 1.00 & 1.00 & 1.00 & 0.00 & 1.00 & 1.00 & 1.00 & 1.00 \\
& R\&D & 1 & 0 &  0.90      & 1.00 & 1.00 & 1.00 & 1.00 & 1.00 & 0.00 & 1.00 & 1.00 & 1.00 & 1.00 \\
& Tool & 7 & 5 &  0.76      & 1.00 & 1.00 & 0.71 & 0.71 & 1.00 & 0.71 & 0.43 & 0.86 & 0.86 & 0.29 \\
& Internal API & 2 & 2 &  0.70      & 0.50 & 1.00 & 0.50 & 0.50 & 1.00 & 1.00 & 0.50 & 1.00 & 1.00 & 0.00 \\
& Internal app & 6 & 6 &  0.55      & 0.83 & 0.50 & 0.67 & 0.67 & 0.50 & 1.00 & 0.00 & 0.67 & 0.67 & 0.00\\\cline{2-15}
& Infrastructure & 3 & 1 &  0.60      & 1.00 & 0.67 & 0.67 & 0.67 & 0.67 & 0.33 & 0.33 & 0.67 & 0.67 & 0.33 \\
& Setup & 2 & 0 &  0.80      & 1.00 & 1.00 & 1.00 & 1.00 & 1.00 & 0.00 & 1.00 & 0.50 & 0.50 & 1.00\\\cline{2-15}
& Documentation & 4 & 0 &  0.40      & 0.50 & 0.75 & 0.50 & 0.50 & 0.50 & 0.00 & 0.75 & 0.00 & 0.00 & 0.50 \\

        \noalign{\smallskip}\hline
    \end{tabular}
    \end{adjustbox}
\end{table}

From this analysis, we conclude that the classifiers trained on Git logs coming from publicly available repositories can be used to identify repositories containing engineered software projects in industrial settings. The results were confirmed by the employees of the companies owning the repositories. However, we identified two limitations of the proposed approach; First, one needs to manually filter repositories being ``forks'' (unless they are supposed to be the subject of the analysis). Second, Git log commit frequencies do not allow distinguishing between the repositories storing application source code and the ones storing documentation written in the markdown language. 

\section{Discussion}
\label{sec:discussion}

PHANTOM extracts five measures in the form of a time series. Then, 42 features are extracted that characterise each of the time series. The measurement and feature extraction process are done directly on the Git log without the need of accessing any external sources of information (e.g. bug trackers, continuous integration servers), analysing project artefacts (e.g. project documentation or source code) or any Git-hosting-site specific information (e.g. GitHub Pull Requests).  This allows PHANTOM to be used to curate Git repositories independently of the infrastructure used to host them. 

Also, the validation shows that PHANTOM was able to rediscover the ground truth of the software repository datasets published in the baseline study by \cite{Munaiah2017}, in which software projects were manually labelled. Many of the PHANTOM (unsupervised) models were able to achieve Precision and Recall close to 1.0.  That confirms that the measures and extracted features are descriptive enough to allow differentiating between engineered and not engineered software projects.

By comparing the prediction accuracy of the PHANTOM models presented in tables \ref{table:accuracy_best_organisation} and \ref{table:accuracy_best_utility} with the accuracy of the classifier in the baseline study presented in Table \ref{table:accuracy_munaiah}, it is clear that the PHANTOM models can compete with the supervised approaches. Committers model achieved slightly higher F-Measure and Recall for the Organization study than the best classifier in the baseline study, and higher Precision for the Utility dataset. This shows, that k-means is a competitive alternative to supervised algorithms for the considered problem. When the features extracted by PHANTOM were used to train supervised models (Random Forest), they obtained similar accuracy than the corresponding models in the baseline study.  

Also, PHANTOM reduces the number of measures needed for analysis from seven taken by \emph{reaper} to one. Although five measurements were experimented with, four of them show competitive results when used on their own. This shows that with only limited information, accurate predictions can be made about the quality of a repository. For instance, PHANTOM can produce competitive results with Commit Frequency alone, which is also taken by \textit{reaper} as part of the seven extracted measures (as a monthly average). This shows that PHANTOM was able to achieve similar accuracy, with a subset of the data used by \emph{reaper} (i.e. one-seventh), by choosing a different representation.

Since measurements are taken from the Git logs, rather than other sources (e.g. source code, GitHub API, GHTorrent), private or closed-source repositories can now be cross-analysed with open-source ones. PHANTOM also avoids the limitations of other sources such as out-of-date information and API key sharing.

PHANTOM’s working assumption is that the programming language of a repository is not relevant. As the Git log is independent of the programming language, it can analyse projects of any programming language, which is a significant improvement over \emph{reaper}. However, the authors must admit that the efficacy of PHANTOM on other programming languages has not been established, since the large dataset contains repositories from a set number of languages, which are the ones supported by \emph{reaper}. 

PHANTOM achieved a 33\% reduction in data collection time over \emph{reaper}, however, 4.64\% of the repositories were unavailable. It took 21.5 days to generate the Git logs for the Large dataset, or one second per repository, which is within 1.3 days of the extrapolated analysis time of 20.2 days. 

PHANTOM reduced the hardware requirements over \textit{reaper} by two orders of magnitude. \textit{Reaper} analysed the Large dataset using a computer cluster of 200 nodes, while PHANTOM achieved the same using a desktop computer.\footnote{Computer hardware and operating system used in this study: Intel i5 CPU, 1Gbps Ethernet, 1TB HDD,  16GB DDR3 RAM, Lubuntu 17.10.} Furthermore, the authors found that the hardware resources were not exhausted (RAM, CPU), spending most of the time idle. The majority of \emph{analysis} time is spent waiting for downloads to complete, rather than Git log extraction. However, the bandwidth was not a limiting factor in the analysis. The bandwidth available to the machine on which PHANTOM ran was 1 Gbps. The authors observed the download speed, which rarely exceeded 40 Mbps. This shows that bandwidth was not the constraint one might expect it to be, but it was rather the speed at which repositories can be downloaded from GitHub. 

Taking into account the reduction in requirements on both processing time and hardware with respect to \emph{reaper} that used a cluster of computers, we can conclude that PHANTOM performs well on commodity hardware, even at large scale.

When both PHANTOM and \emph{reaper} are applied to the Large dataset (unlabelled data), the best models obtained similar accuracy ca. 0.70. When PHANTOM and \emph{reaper} were unanimous in their predictions, they were usually correct. We also observed that there were ca. 6\% of instances for which the models provided strongly contradict predictions and have the accuracy similar to random guessing. Finally, the best-performing models provided similar predictions of the number of engineered projects hosted on GitHub, being 19\% and 24\%. However, the remaining PHANTOM models, that were only slightly less accurate than the best-performing model, estimated this ratio to 38--55\%. 


To summarise the discussion, we  present a side-by-side comparison between PHANTOM and \emph{reaper} in Table \ref{table:comparison}. 

\begin{table}
    \centering
       \caption{Comparison between \textit{reaper} and PHANTOM.}
    \label{table:comparison}
    \begin{tabularx}{\textwidth}{l | c c}
    \hline\noalign{\smallskip}
        Aspect  & \textit{reaper}   & PHANTOM          \\ 
        \noalign{\smallskip}\hline\noalign{\smallskip}
        Data Collection Time & \textgreater{}1 month     & 3 weeks          \\ 
        Hardware Requirements   & Computer cluster (200 nodes) & Desktop computer \\ 
        Measures    & 7 measures& 1 measure        \\ 
        Data Sources  & GHTorrent, source code      & Git              \\ 
        Sample Size   & 1,857,423      & 1,771,167        \\ 
        Machine Learning approach   & Supervised   & Unsupervised     \\ 
        F-Measure (Organization)   & 68\%      & 77\%             \\ 
        F-Measure (Utility)      & 84\%        & 76\%             \\ 
        Percentage engineered      & 24\%      & 19\%          \\ 
        Programming Languages Supported& 
        C, C\#, C++, Java, PHP, & Any \\ 
        &  Python, Ruby& \\ 
        Implementation Languages   & Python      & Python, Rust\\
        \noalign{\smallskip}\hline
    \end{tabularx}
 
\end{table}

We applied PHANTOM to curate 100 repositories owned by two companies and learned that the models calibrated on the open-source datasets were able to correctly identify repositories containing projects developed for customers. Unfortunately, we have also observed that the method is not able to recognise the repositories being ``forks'' or to distinguish between the code of software applications and the documentation written using the markdown language. Still, we did not find any repositories that would contain only documentation in the analysed sample of 250 repositories from the Large dataset. Therefore, the impact of the latter issue on the results of our study could be considered as negligible.

Based on the results, we cannot firmly state which of the measures provide the best capabilities of filtering engineered projects. Therefore, it is worth considering using ensembles of PHANTOM models (e.g. use the majority voting). We also recommend setting the correlation threshold between 0.75 and 0.90 since the resulting models should have very similar accuracy.

\subsection{Threats to Validity}

Since we mostly base our study on the datasets published by the baseline study, we are exposed to similar threats to validity.

\paragraph{Constructs Validity}  The general definition of ``engineered software project'' refers to practices leveraged by a software project while the customised definition proposed by \citeauthor{Munaiah2017}is based on the notion of similarity between software repositories. Although \citeauthor{Munaiah2017} were investigating the quality of project artefacts in seven dimensions while constructing the reference Organization and Utility datasets, it does not prove the equivalence of both definitions.

In our study, we use five measures (Integration Frequency, Commit Frequency, Integrator Frequency, Committer Frequency, and Merge Frequency) that are taken from Git logs and have the form of time series (measurement performed in time). Therefore, they could reflect the ways of working in the projects, also revealing some periodic behaviour. However, none of these measures can be used to evaluate the quality of artefacts  (e.g. quality of code or software documentation).

\paragraph{Internal Validity} These five measures we use in the study have been selected by the authors and may not provide full characteristics of the repositories. In addition, feature vectors contain 42 features, which are also chosen by the researchers. Although attention has been paid to choose features that are reflective of the time series, no rigorous process was followed to ensure they were so.  

The procedure of determining the ground truth for the datasets published in the baseline study by \citeauthor{Munaiah2017} and for the Industry dataset differ. For the latter dataset, we base it on the opinion of the companies' employees. Although we believe that the employees had enough information to judge whether the predictions made by PHANTOM were correct (i.e. having direct access to the code and other project's artefacts, knowledge about the projects and processes used in the companies), we interviewed only one representative from each company to verify the predictions. Finally, we used different versions of the PHANTOM tool to analyze the datasets published by \citeauthor{Munaiah2017} and the Industry dataset. Therefore, there could be some differences in the feature-extraction algorithm between the versions that could influence the results. However, since we use different feature sets, the impact of these differences should be minor and negligible.

\paragraph{External Validity} The ground truth of the dataset published by \cite{Munaiah2017}, which is based on a description of 300 engineered and 150 not engineered repositories may not agree with other collections of repositories. Although the authors cannot confirm the correctness of the ground truth, PHANTOM uses unsupervised models that were able to rediscover the ground truth by using only the features space. The produced clusters agree with the ground truth to a large degree, which supports its correctness. That being said, the possibility also exists that both PHANTOM and the ground truth are wrong and this agreement is coincidental. However, the additional studies performed on a new sample of repositories from the Large dataset and the study performed on the Industry dataset showed that the method could be used in different contexts.

Also, we have to accept the fact that the statements about the download speed may not be relevant to researchers with different hardware, Internet connection, or an alternative agreement with GitHub.

Finally, the predicted number of engineered projects in the Large dataset differs between PHANTOM models. A similar observation was made in the baseline study, for which the predicted percentage of engineered projects for the Large dataset ranged between 6\% and 70\%. However, the predictions made by two best-performing models in both studies provided similar estimates.
Nevertheless, we believe that the predictions made by both studies should be taken with caution since the level of uncertainty is high.

\subsection{Ethical Considerations}

The most important ethical consideration concerns the Git logs published as part of this paper, which contain the names and emails of GitHub users. Although these are publicly available, the authors have anonymised this data to protect the users’ privacy, in accordance with GitHub terms and conditions. Therefore, the published Git logs contain placeholder names and emails which neither hinder analysis nor leak sensitive information.

A further ethical consideration is the collection of repositories from GitHub. Although the repositories are publicly available, mining data on the scale seen in this paper is not generally acceptable behaviour according to GitHub’s terms of service \cite{GitHubToS}. The authors came to an agreement with GitHub about the duration and use of GitHub’s servers, and how the collection should be carried out. GitHub has requested that the details of this agreement should not be published, because it is specific between GitHub and the authors. It is important to emphasise that contacting GitHub before mining is a necessity for ethical research, due to the terms of service. This extends (beyond cloning from GitHub) to using the GitHub API.

\section{Conclusions}
\label{sec:conclusions}

The amount of available software repositories rises everyday; between January 2014 and March 2018 the number of GitHub repositories rose from 10.6 million to over 80 million, an increase of 780\%. This dramatic increase has not been matched by analysis methods so far. In order to make use of this large corpus of data, it is essential to filter out undesirable repositories, however as \cite{Munaiah2017} puts it: \emph{``there are limited means of separating the signal (e.g. repositories containing engineered software projects) from the noise (e.g. repositories containing homework assignments).''}

The problem with the majority of applied filtering techniques when mining software repositories is that they are inaccurate or unproven. \cite{Munaiah2017} introduced a new method(\textit{reaper}) that achieves high accuracy, but requires the computing power of a computer cluster, and cannot be applied to all repositories as it is based on static analysis of code. 

In this study, we proposed a new method called PHANTOM that improves on existing methods with respect to analysis time and hardware requirements, without sacrificing accuracy. Therefore, the barrier for researchers, who often do not have access to expensive hardware, is removed. PHANTOM achieves this by using a time-series representation as input to create feature vectors describing properties of the time-series (e.g. the number of peaks). These time-series are based on information from the development history and are transformed to feature vectors that can be used with machine learning algorithms for smarter comparison. In particular, this makes the time-series compatible with a standard k-means algorithm, which is used to cluster the repositories into two groups; engineered and not engineered. As a result, PHANTOM allows researchers to automatically and inexpensively filter large datasets of unknown quality and remove repositories that are undesirable for further more specific analysis.

In the performed validation, PHANTOM was able to rediscover a ground truth of 450 repositories, with the best k-means models achieving up to 1.0 Precision and Recall. When applied to new, unseen data, the best models in PHANTOM achieved up to 0.87 Precision or 0.94 Recall. The MCC of the best models was overall positive, with the highest being 0.65. This is competitive to the best supervised classifiers from the baseline study by \cite{Munaiah2017}  that reported 0.88 Precision and 0.99 Recall on the same datasets. PHANTOM obtained the metadata of 1,786,601 GitHub repositories in 21.5 days, which is over 33\% faster than \emph{reaper}, and reduced the hardware requirements by two orders of magnitude. The best-performing model predicted that 19\% of the analysed GitHub repositories contain engineered projects, compared to 24\% reported in the baseline study. Finally, we applied PHANTOM to curate 100 repositories owned by two companies and learned that the model is capable of recognising external products as engineered projects. However, we also identified some limitations of the method in correctly identifying forked repositories and repositories containing documentation as engineered software projects.

Because of the limitations of existing curating methods, many studies mining software repositories use inaccurate or unproven filtering approaches (e.g. popularity). PHANTOM could be applied in such studies to improve the data curation process. For example, \cite{Robles2017} published a collection of 24,000 repositories, for which PHANTOM could be useful to filter out undesirable repositories. Such use cases are possible because PHANTOM is not dependent on mirroring services like GHTorrent. Any Git repository, not just those available through such services, can be analysed, making PHANTOM also suitable for research on private, or very specific collections of repositories. Furthermore, PHANTOM is programming language agnostic.

\begin{acknowledgements}

This research has been supported by Software Center (www.software-center.se), Chalmers $|$ University of Gothenburg, and the statutory funds of Poznan University of Technology.

\end{acknowledgements}


%
%

\end{document}